\documentclass[journal,doublecolumn]{IEEEtran}
\usepackage{amsmath,amssymb,placeins,makecell,stfloats,cite}
\usepackage{placeins}
\usepackage{mathrsfs}
\usepackage{array}
\usepackage[dvips]{graphicx}
\usepackage[usenames]{color}
\usepackage{amsfonts}
\usepackage{latexsym}
\usepackage{subfigure}
\usepackage[justification=centering]{caption}
\usepackage{floatrow}
\usepackage[format=hang]{subfig}
\newtheorem{theorem}{{{\textit{Theorem}}}}

\newtheorem{lemma}{{{\textit{Lemma}}}}
\newtheorem{corollary}{{{{\textit{Corollary}}}}}

\newtheorem{definition}{{{\textit{Definition}}}}

\newtheorem{remark}{{{\textit{Remark}}}}

\newtheorem{example}{{{\textit{Example}}}}

\begin{document}
\title{A Direct Construction of Optimal ZCCS With Maximum Column Sequence PMEPR Two for MC-CDMA System }
\author{Palash~Sarkar,~%\IEEEmembership{}
        and Sudhan~Majhi%\IEEEmembership{Senior Member,~IEEE,}
        %and~Zilong~Liu%,~\IEEEmembership{Member,~IEEE}% <-this % stops a space
\thanks{Palash Sarkar is with Department of Mathematics and Sudhan Majhi is with the Department of Electrical Engineering, Indian Institute of Technology Patna, India, e-mail: {\tt palash.pma15@iitp.ac.in; smajhi@iitp.ac.in}.}}
  %      and~Zilong~Liu%,~\IEEEmembership{Member,~IEEE}% <-this % stops a space
%\thanks{This  work  was  supported  in  part  by  the  Visvesvaraya Young Faculty Research Fellowship, Ministry of Electronics  and Information Technology,  
%Government  of  India,  being  implemented  by  the  Digital  India Corporation  and  Early  Career  Young  Scientists  by  the  Science  and  Engineering  
%Research  Board  under  the  Department  of  Science  and  Technology, Government  of  India.
%
%Palash Sarkar is with Department of Mathematics and Sudhan Majhi is with the Department of Electrical Engineering, Indian Institute of Technology Patna, India, e-mail: {\tt palash.pma15@iitp.ac.in; smajhi@iitp.ac.in}.}% <-this % stops a space
%\thanks{Zilong Liu is with Institute for Communication Systems, 5G Innovation Centre, University of Surrey, UK, e-mail: {\tt zilong.liu@surrey.ac.uk}.}}% <-this % stops a space
\IEEEpeerreviewmaketitle
\maketitle
\begin{abstract}
%Multicarrier Code Division Multiple Access (MC-CDMA) scheme is the promising candidate for future
%broadband wireless systems, as it provides higher flexibility, transmission rates and spectral efficiency. It
Multicarrier code-division multiple-access (MC-CDMA) combines an orthogonal frequency division multiplexing (OFDM) modulation and a code-division
multiple-access (CDMA) scheme to exploits the benefits of both the technologies. The high peak-to-mean envelope power ratio (PMEPR) is a considerable
problem in MC-CDMA system. However, the problem can be addressed by utilizing complete complementary codes (CCCs) in MC-CDMA system. But the set size upper bound of CCC does not allow the system
to support large number of users for a given number of subcarriers in the system.
%However, MC-CDMA system has
%high peak-to-mean envelope power ratio (PMEPR) which can be dealt by employing complete complementary codes (CCCs). 
%This problem can be dealt by employing proper codes. A
%Z-complementary code set (ZCCS) is a collection of codes with same length and equal number of constituent sequences or two 
%dimensional matrices with zero correlation zone (ZCZ) properties. 
In a CCC and
Z-complementary code set (ZCCS) based asynchronous MC-CDMA system, the PMEPR is determined by column sequence PMEPR of the codes. 
%However, the main drawback of CCC is that the set 
%size is upper bounded by the number of constituent sequences. 
In order to suuport a large number of users with low column sequence PMEPR, in
this paper, we have proposed a new optimal ZCCS with larger set 
size.
%to reduce column sequence PMEPR of quasy asynchronous MC-CDMA. 
The code is constructed using Boolean function approach, i.e., by a 
direct construction method. The number of constituent sequences in ZCCS is the same as the number of subcarriers in MC-CDMA. So, large size ZCCS 
for large number of users in MC-CDMA can be constructed through a rapid hardware generation. The proposed ZCCS  
has mximum column sequence PMEPR of 2 and it achieves the theoretical upper bound of optimality. 
%The column sequence PMEPR of existing 
%ZCCS by using Boolean function approach is determined by the number of constituent sequences which is same as the number of sub-carrier
%in a ZCCS based MC-CDMA system. 
%The number of sub-carrier in a MC-CDMA system is 
%generally large which degrades the performance of a ZCCS based MC-CDMA system, although Boolean function approach is a direct method to construct
%ZCCS and it is feasible for rapid hardware generation particularly
%for long ZCCSs.
%Although, all existing Boolean function approach is a direct method to construct
%ZCCS and it is feasible for rapid hardware generation particularly
%for long ZCCSs, the large number of sub-carrier in a MC-CDMA system degrades the performance of a existing ZCCS based MC-CDMA system.   
%In this paper, we propose a construction of ZCCS with maximum column sequence PMEPR $2$ by using generalized 
%Boolean functions (GBFs). 
%The  optimal with respect to the theoritical upper bound. 
Our proposed construction can
also generate inter-group complementary (IGC) code set for MC-CDMA with the same PMEPR. 
%whereas the maximum column sequence PMEPR of all existing IGC code set is given by the number of constituent sequnces in a code. 
This work also establishes a link from ZCCS and IGC code set to higher-order ($\geq 2$) Reed-Muller (RM) code.  
\end{abstract}
%\IEEEpeerreviewmaketitle
\section{Introduction}\label{sec:intro}
Multicarrier code-division multiple-access (MC-CDMA) is a most promising technology for fifth 
generation (5G) and beyond wireless communication. It has brilliant features as it combines multicarrier modulation and multiplexing technique. 
However, it suffers from high peak-to-mean envelope power ratio
(PMEPR) problem. 
%Spreading sequences play an important role in code-division multiple-access (CDMA) to determine it's system performance.
%In code-division multiple-access (CDMA) technique, one of the important spread-spectrum approach is multicarrier CDMA (MC-CDMA) where 
%the multicarrier communication signals suffer due to high peak-to-mean envelope power ratio (PMEPR) problem. 
The high PMEPR value problem in 
MC-CDMA system can be compensated by employing proper spreading codes  which provide low column sequence PMEPR. Because of suitable auto- and cross-correlation properties, such spreading codes 
\cite{arthina,cccsmajhi,p_tcom,chen_zccs,li_ZCCS,das_sp,avik_elect,liano} are also used to deal multiple access interference (MAI) and multipath interference (MPI) besides 
PMEPR problem.
 In this context, we introduce Golay complementary pair (GCP),
complete complementary code (CCC), and ZCCS. A pair of sequences, with the sum of their aperiodic auto-correlation function (AACF) to zero for all nonzero time shift, called
GCP \cite{gol1961}. Sequences of a GCP is known as Golay sequences. The concept
of complementary code (CC) was introduced by Tseng and Liu
in \cite{chong1972} by extending the idea of GCP. The sum of AACFs of the sequences in a CC become zero for all out of phase shift. 
In 1999, Davis \emph{et~al.} proposed a constrcution of GCP in \cite{Davis1999}, known as Golay-Davis-Jedweb (GDJ), by using second-order 
generalized Boolean function (GBF) and provided a link between their GCPs and
Reed-Muller (RM) code. 
%The sequences obtained by Davis \emph{et~al. } in \cite{Davis1999} are called Golay-Davis-Jedweb (GDJ) sequences in this paper.
Later,  Paterson \emph{et~al.} proffered a construction of CC 
%in \cite{pater2000} by associating it with 
by using graph and second-order 
RM code in \cite{pater2000} and the work is generalized by Schmidth by using higher-order RM in \cite{kusch}. 
%Paterson's work was further generalized by Schmidth in \cite{kusch} by where a established a link between CC with 
%higher-order ($\geq 2$) RM code. 
Paterson's idea of CCs were extended to CCC by Rathinakumar \emph{et~al.} in \cite{arthina} by using second-order
GBF.
%and also established a link between second-order RM code and CCC. 
A set of CCs with ideal cross-correlation properties is said to be CCC if the number of CCs is equal to
the number sequences in each CC.
%, is called CCC if the 
%CCs have ideal auto and cross-correlation properties. 
A construction of CCC were introduced in \cite{uda2014} where
it was shown that the column sequence PMEPR of a CCC based MC-CDMA system can have at most $2$ unlike the CCC 
introduced in \cite{arthina}.

ZCCS has the same correlation properties as CCC   
%s collection of two dimensional matrices (or codes) of equal order with ideal aperiodic auto and cross-correlation properties inside
inside a zone, called zero correlation zone (ZCZ). As compared with CCC, ZCCS has much larger set size \cite{liu2011} which allows a ZCCS based MC-CDMA system
to support a large number of users unlike CCC based MC-CDMA system where number of subcarriers
%\footnote{In CCC based MC-CDMA system number of sub-carriers is equal to number of users.} 
is equal to the number of users. Having the ZCZ properties, ZCCS is used to mitigate
MAI for received multiuser quasi-synchronous signals within the ZCZ width \cite{Liu-ITW2014}.
In 2007, Fan \emph{et~al.} \cite{fan2007} introduced binary ZCCS and it is generalized to pairwise ZCCS by 
Feng \emph{et~al.} \cite{lfengispl2008} in 2008. In 2019, a construction of ZCCS has been 
introduced by Palash \emph{et~al.} in \cite{p_tcom} by associating it with second-order RM code and graph. Another construction 
of ZCCS has been reported in \cite{chen_zccs} by using second-order GBFs. In 2015, a construction of ZCCS which has maximum column sequence
PMEPR of $2$, was introduced by Li \emph{et~al.} in \cite{li_ZCCS}. The construction is based on  Golay sequences 
and orthogonal matrix. But this construction is not a direct construction and it may not be advantageous for the hardware generation of long ZCCSs. To reduce high 
PMEPR problem and in order to support a large number of users in a MC-CDMA system, the aim of this paper is to 
provide a direct construction of a new ZCCS based on which a MC-CDMA system can have PMEPR of at most $2$.

A ZCCS is known as inter-group complementary (IGC) code set when it is divided into numerous distinct code groups with the properties
that the AACF of 
each code is ideal within the ZCZ width. The aperiodic cross-correlation function (ACCF) of two disjoint codes drawn from the same code group is also ideal inside the ZCZ
width. The ACCF of two codes drawn from two different code groups is zero for all time shifts. In 2008, Li \emph{et~al.}
proposed a construction of IGC code set based on CCCs in \cite{jli_igc_2008}. Their code assignment algorithm shows that the CDMA systems employing the
IGC codes (IGC-CDMA) outperform traditional CDMA with
respect to bit error rate (BER). The ZCZ width of IGC code set in \cite{jli_igc_2008} depends on the length of constituent sequences of 
CCs and the construction is not direct. Recently, a direct construction of IGC code set has introduced in
\cite{sarkar_igc} by using second-order GBFs. However, the constructions given in \cite{jli_igc_2008,sarkar_igc} cannot provide a tight 
column sequence PMEPR as the column sequnce PMEPRs of IGC code sets from both of the constructions is upper bounded by the number
of constituent sequences which motivate us to provide a direct constrcution of IGC code set with maximum column sequence PMEPR $2$. 

In this paper, we first propose a direct construction of ZCCS  by using higher-order ($\geq 2$)
GBFs. The maximum column sequence PMEPR of our proposed ZCCS based MC-CDMA system is $2$ unlike the ZCCS given in \cite{p_tcom,chen_zccs}. Then we show 
that our propose ZCCS can also generate IGC code set with maximum column sequence PMEPR of IGC based MC-CDMA is $2$ which make our construction more 
efficient than existing IGC code set construction. Our propose construction establish a relation of ZCCS and IGC code set with higher order ($\geq 2$) RM code.
%between ZCCS and higher-order
%($\geq 2$) RM code and also a relation between IGC code set and higher-order RM code. 
We also relate our constructions with graph. Specially,
we have shown that our propose construction generates ZCCS corresponding to a GBF if the graphs of all possible restrictions of the GBF over 
some fixed specific variables, contain a path and some fixed isolated vertices. The construction generates IGC code set if the GBF does not contain
a term which is associated with the restricted variables and the variables which appear as isolated vertices in the grpahs of restricted Boolean
functions.

The paper is arranged as follows. In Section II, some definitions and useful notations are presented. 
A construction of ZCCS with maximum column sequence PMEPR $2$ has been presented in Section III. In Section IV, a construction IGC code set with maximum
column sequence PMEPR 2 is presented. We compare our proposed construction with existing construction in Section V.
Finally, we conclude our proposed constrcution in Section VI.
\section{Preliminary}
\label{sec:back}
\subsection{Definitions of Correlations and Sequences}
Let $\textbf{a}=(a_0,a_1,\hdots, a_{L-1})$ and $\textbf{b}=(b_0,b_1,\hdots, b_{L-1})$ be two complex-valued sequences of equal length $L$. For an integer $\tau$, define
\begin{equation}\label{equ:cross}
\mathscr{C}(\textbf{a}, \textbf{b})(\tau)=\begin{cases}
\sum_{i=0}^{L-1-\tau}a_{i+\tau}b^{*}_i, & 0 \leq \tau < L, \\
\sum_{l=0}^{L+\tau -1} a_ib^{*}_{i-\tau}, & -L< \tau < 0,  \\
0, & \text{otherwise},
\end{cases}
\end{equation}
and $\mathscr{A}(\textbf{b})(\tau)=\mathscr{C}(\textbf{b},\textbf{b})(\tau)$.
The following functions $\mathscr{C}(\textbf{a}, \textbf{b})$ and $\mathscr{A}(\textbf{b})$ are called ACCF of $\textbf{a}$ and $\textbf{b}$, and
AACF of $\textbf{b}$ respectively.
%These functions are called ACCF
%between $\textbf{a}$ and $\textbf{b}$ and the AACF of $\textbf{b}$, respectively.
%We also define correlation functions for $\mathbb{Z}_q$ valued vector, where $q(\geq 2)$ is an even number.
%We do this by defining $\omega=e^{2\pi i/q}$ and associating it
%with each element of the complex vector $\textbf{a}=(\omega^{a_0},\omega^{a_1},\cdots, \omega^{a_{n-1}})$, where $a_i\in \mathbb{Z}_q$.
%\begin{definition}
%A set of $M$ sequences  $\textbf{a}_0,\textbf{a}_1, \cdots ,\textbf{a}_{M-1}$, each  is said to be a complementary set if \\
%$ A(\textbf{a}^0)(\tau)+A(\textbf{a}^1)(\tau)+\cdots +A(\textbf{a}^{N-1})(\tau)=0 $, $\tau\neq 0$.
%\end{definition}
%Notice that the definition applies equally to $\mathbb{Z}_q$-valued and complex-valued vectors.
%A complementary set of size 2 is called GCP.
Let $\textbf{C}=\{C_0,C_1, \hdots ,C_{K-1}\}$ where
\begin{equation}
\begin{split}
C_\mu=\begin{bmatrix}
\textbf{a}_0^\mu \\ \textbf{a}_1^\mu \\ \vdots \\ \textbf{a}_{M-1}^{\mu}
\end{bmatrix}_{M\times L}
=\left[\textbf{d}_0^{\mu}~\textbf{d}_1^{\mu}~\cdots~\textbf{d}_{L-1}^{\mu}\right],
\end{split}
\end{equation}
where $\textbf{a}_p^\mu$ ($0\leq p \leq M-1,0 \leq \mu \leq K-1$) is the $p$th row sequence or $p$th constituent sequence  of $C_\mu$
and $\textbf{d}_{e}^\mu$ ($0\leq e \leq L-1$) is the $e$th column sequence of $C_\mu$.
For $C_{\mu}$, $C_{\nu}\in \textbf{C}$ $(0\leq \mu,\nu\leq K-1)$, the ACCF
of $C_{\mu}$ and $C_{\nu}$ is defined by
 \begin{equation}
  \mathscr{C}(C_{\mu},C_{\nu})(\tau)=\displaystyle \sum_{p=0}^{M-1}\mathscr{C}(\textbf{a}_p^{\mu},\textbf{a}_p^{\nu})(\tau).
 \end{equation}
\begin{definition}\label{def_ccc}
$\textbf{C}$ is called  CCC if $K=M$ and it satisfies the following properties:
 \begin{equation}
  \mathscr{C}(C_{\mu},C_{\nu})(\tau)
=\begin{cases}
LM, & \tau=0, \mu=\nu;\\
0, & 0<|\tau|<L, \mu=\nu;\\
0, & |\tau|< L, \mu\neq \nu.
\end{cases}
 \end{equation}
\end{definition}
The code $C_\mu$ $(0\leq \mu \leq K-1)$, is called CC and it is called GCP if it contains a pair of sequences.
%When $M=2$, $C_\mu$
%reduces to a GCP and either sequence of the pair is called a Golay sequence.
\begin{definition}
$\textbf{C}$ is said to be ZCCS and we denote it by $(K,Z)$-$\text{ZCCS}_M^L$ if it satisfies the following properties:
\begin{eqnarray}
\mathscr{C}(C_{\mu},C_{\nu})(\tau)
=\begin{cases}
LM, & \tau=0, \mu=\nu,\\
0, & 0<|\tau|<Z, \mu=\nu,\\
0, & |\tau|< Z, \mu\neq \nu,
\end{cases}
\end{eqnarray}
 where $Z$ is called ZCZ width.
 \end{definition}
 \begin{definition}\label{defzccs}
 Let $\textbf{C}$ can be expressed as the union of $M$ distinct code groups $\mathcal{I}_g$ $(g=0,1,\hdots, M-1)$ where each code group
 is a collection of $K/M$ codes and $K=ML/Z$. $\textbf{C}$ is said to be IGC code set and denoted by $\mathcal{I}\left(K,M,L,Z\right)$ if it
 satisfies the following properties:
 %Given an IGC code set $\mathcal{I}\left(K,M,L,Z\right)$  \cite{jli_igc_2008} 
	%where $K$ denotes number of codes, $P$ denotes number of constituent sequences in each
	%code, $L$ denotes the length of each constituent sequence and $Z$ denotes ZCZ width
%	, where {\color{blue}$K=ML/Z$}. The $K$ codes can be divided into 
%	$M$ code groups denoted by $\mathcal{I}_g$ $(g=0,1,\hdots, M-1)$, each group contains $K/M=L/Z$ codes. The code set $\mathcal{I}\left(K,M,L,Z\right)$ has the following properties:
	\begin{equation}
	\mathscr{C}\!\left(C_\mu,C_\nu\right)\!\!\left(\tau\right)\!=\!\begin{cases}
	ML, & \tau=0,\mu=\nu,\\
	0, & 0<|\tau|<Z,\mu=\nu,\\
	0, & |\tau|<Z,\mu\neq \nu, C_\mu,C_\nu\in \mathcal{I}_g,\\
	0, & |\tau|\!\!<\!\!L,C_\mu\!\!\in\!\! \mathcal{I}_{g_1}, C_\nu\!\!\in\!\! \mathcal{I}_{g_2}, g_1\!\!\neq\!\! g_2,\\
	\textnormal{others}, & \textnormal{otherwise}.
	\end{cases}
	\end{equation}
\end{definition}
 \subsection{Peak-to-Mean Envelope Power Ratio (PMEPR)}
%For $q$-PSK modulation, the OFDM signal for the word 
Let $\textbf{A}=(A_0,A_1,\hdots, A_{M-1})$ be a complex valued sequence of length $M$. For 
a multi-carrier system with $M$ subcarriers, the time domain multi-carrier signal can be written as \cite{uda2014}

%(where $a_i\in \mathbb{Z}_q$) can be modeled as the real part
%of 
\begin{equation}
 S(\textbf{A})(t)=\displaystyle\sum_{j=0}^{L-1}A_{j} e^{2\pi \sqrt{-1}jt},
\end{equation}
where the carrier spacing has been normalized to $1$ and $\textbf{A}$ is spreaded over $M$ subcarriers. 
%$\omega_q=\exp(2\pi \sqrt{-1}/q)$ is a complex $q$th root of unity and $f_0+jf_s$ $(0\leq j< L)$ is $j$th carrier frequency of the OFDM signal.
%We define the instantaneous envelope power of the OFDM signal as \cite{pater2000}
 %\begin{equation} \nonumber
Denote  $P(\textbf{a})(t)=|S(\textbf{a})(t)|^2$.
 %\end{equation}
% From the above expression, it is easy to derive that
% \begin{equation}
% \begin{split}
%  P(\textbf{a})(t)&=\displaystyle\sum_{\tau=1-L}^{L-1}A(\textbf{a})(\tau)\exp(2\pi \sqrt{-1}\tau f_s t)\\
%  &=A(\textbf{a})(0)+2\cdot \text{Re}\left\{ \displaystyle\sum_{\tau=1}^{L-1}A(\textbf{a})(\tau)\exp(2\pi \sqrt{-1}\ell f_s t)\right\},
%\end{split}
%\end{equation}
%where $\text{Re}\{x\}$ denotes the real part of a complex 
%number $x$. 
The PMEPR of a polyphase sequence
$\textbf{A}$ under the multi-carrier modulation is defined as
\begin{equation}\nonumber
\textnormal{PMEPR}(\textbf{a})= \frac{1}{M}\displaystyle \sup_{0\leq  t<1}P(\textbf{A})(t).
\end{equation}
%The largest value that the PMEPR of an $n$-subcarrier OFDM signal is $n$.

Let $C_\mu$ be a code from the ZCCS $\textbf{C}$ which is defined in \textit{Definition \ref{defzccs}}. In a ZCCS based
MC-CDMA system, $\textbf{a}_\nu^\mu$ is spread in $\mu$th subcarrier over $L$ chip-slots and $\textbf{d}_{\nu'}^\mu$  is spread in the
$\nu'$th chip-slot over $M$ subcarriers. The PMEPR of $C_\mu$ is given by
\begin{equation}
 \textnormal{PMEPR}\left(C_\mu\right)=\sup_{0\leq \nu'<L}~\textnormal{PMEPR}\left(\textbf{d}_{\nu'}^\mu\right).
\end{equation}
A CCC based MC-CDMA system transmitter structure is given by Liu \emph{et~al. } in \cite{uda2014}. 
A ZCCS based MC-CDMA is given in \cite{li_ZCCS} and QCSS based MC-CDMA is given in \cite{abdus}. 
%From the theoretical bound, for a $(K,Z)ACS_M^N$, there is $K\leq M\lfloor N/Z\rfloor$. If the equality holds then $(K,Z)ACS_M^N$ is an optimal.
\subsection{Generalized Boolean Functions and Graphs}
There are $2^m$ distinct monomials which are of degree
$0,1,\hdots,m$ over the variables $x_0,x_1,\hdots,x_{m-1}$. If $\mathcal{S}_r$ is the set of all monomials of degree at most $r$, $\mathcal{S}_r$
can be expressed as 
\begin{equation}
\begin{split}
 \mathcal{S}_r&=\left\{x_{\alpha_1}x_{\alpha_2}\cdots x_{\alpha_k}:0\leq  k\leq r,\right. \\&
 ~~~~~~~~~~~\left.0\leq\alpha_1<\alpha_2<\cdots<\alpha_k\leq m-1 \right\},
\end{split}
 \end{equation}
where $\mathcal{S}_r$ contains $\displaystyle\sum_{\alpha=0}^r$\!${m}\choose{\alpha}$ distinct monomials of degree $0$ to $r$ ($0\leq r\leq m$). A $r$th degree GBF $f$ of 
$m$ variables $x_0,x_1,\hdots,x_{m-1}$ over $\mathbb{Z}_q$ can uniquely be expressed as a linear combination of monomials from the set
$\mathcal{S}_r$ with $\mathbb{Z}_q$-valued coefficients provided that the coefficient of at least one of the $r$th order monomials is nonzero.
%
%A monomial of degree $k$ is defined as the product
%of any $k$ distinct variables among  $x_0,x_1\cdots x_{m-1}$. There are $2^m$ distinct monomials over $m$ variables listed below:
%\begin{equation}
%\begin{split}
%1,x_0,x_1,\cdots,x_{m-1},x_0x_1,x_0x_2,\cdots,x_{m-2}x_{m-1},\cdots,\\x_0x_1\cdots x_{m-1}.
%\end{split}
%\end{equation}
%A function $f$ is said to be a GBF if it can uniquely be expressed as a linear
%combination of these $2^m$ monomials, where the coefficient of each monomial is drawn from $\mathbb{Z}_q$.
For a second-order GBF $f$, the graph of $f$ is denoted by $G(f)$ which contains a edge between the vertices $x_i$ and $x_j$ if there
is a term $q_{i,j}x_ix_j$ $(0\leq i<j\leq m-1,q_{i,j}\neq 0 )$ in the expression of $f$.
%{\color{blue}denotes} the graph of $f$ which is obtained by joining {\color{blue}between} the vertices $x_i$ and $x_j$ by an edge if there is
%a term $q_{i,j}x_ix_j$ $(0\leq i<j\leq m-1)$ in the GBF $f$ with $q_{i,j}\neq 0$ $(q_{i,j}\in \mathbb{Z}_q)$.  
The complex-valued sequence corresponding to $f$ is expressed as follows:
\begin{equation}
 \psi(f)=(\omega^{f_0}, \omega^{f_1}, \hdots, \omega^{f_{2^m-1}}),
\end{equation}
where $f_i=f(i_0,i_1,\hdots,i_{m-1})$, $\omega=\exp(2\pi\sqrt{-1}/q)$, $q$ ($\geq 2$) is an even number, and $(i_0,i_1,\hdots,i_{m-1})$ is the 
binary vector representation of $i$. Below some notations are presented for better presentation of the paper:
\begin{itemize}
 \item $\tilde{f}$ denotes $f(1-x_0,1-x_1,\hdots,1-x_{m-1})$.
 \item $\bar{x}$ denotes the binary complement of $x\in \{0,1\}$.
 \item $\textbf{a}^*$ is the complex conjugate of a complex-valued vector $\textbf{a}$.
 \item $J=\{j_0,j_1,\hdots,j_{k-1}\}$ ($\subset \{0,1,\hdots,m-1\}$).
 \item $\textbf{x}_J=(x_{j_0},x_{j_1},\hdots, x_{j_{k-1}})$.
 \item $\textbf{c}=(c_0,c_1,\hdots,c_{k-1})\in\{0,1\}^k$.
\end{itemize}
Consider the function $f\arrowvert_{x_j=c}$, obtained
by substituting $x_j=c$ in $f$, be equivalent to the graph
obtained by deleting the vertex $x_j$ and all the edges associated with $x_j$ from $G(f)$. Similarly,  $G(f\arrowvert_{\textbf{x}_J=\textbf{c}})$ is obtained by
deleting the vertices $x_{j_0},x_{j_1},\hdots,x_{j_{k-1}}$ from $G(f)$. The $i$th component of the complex-valued sequence 
$\psi(f\arrowvert_{\textbf{x}=\textbf{c}})$ is denoted by
$\omega^{f(i_0,i_1,\hdots, i_{m-1})}$ if $i_{j_\alpha}=c_{\alpha}$ for each
$0\leq \alpha <k$ and equal to zero otherwise. 
%For more details, readers can go through \cite{p_tcom}.
 
% Let $Q$ be the quadratic form of $f$. Then, the GBF $f$ can be expressed as 
A second-order GBF $f$ can be expressed as 
\begin{equation}\label{Bool}
f=Q+\displaystyle \sum_{i=0}^{m-1}g_ix_i+g',
\end{equation}
where $Q$ is the quadratic form present in $f$ and $g',g_i\in \mathbb{Z}_q$. For more details, readers can go through \cite{p_tcom,arthina}.
%The final graph is independent of the choice of $\textbf{c}$. That is, for any
%$\textbf{c}$, the quadratic part of the function $f\arrowvert_{\textbf{x}=\textbf{c}}$ is completely described by
%the graph which is obtained from $G(f)$ by deleting vertices $x_{j_0},x_{j_1},\hdots,x_{j_{k-1}}$.
%Note that the quadratic forms in the
%functions $f$ and $\tilde{f}$  are the same and therefore, they have the same associated graph.
%Corresponding to
%each GBF $f$, we define a complex valued sequence $\psi(f)$ of length $2^m$ by defining
%$\psi(f)=(\omega^{f_0}, \omega^{f_1}, \hdots, \omega^{f_{2^m-1}})$, where $f_i=f(i_0,i_1,\hdots,i_{m-1})$, $\omega=\exp(2\pi\sqrt{-1}/q)$ $(q$ is a positive integer no less than  $2)$
%and $(i_0,i_1,\hdots,i_{m-1})$ is the binary vector representation of integer $i$ $(i=\displaystyle \sum_{j=0}^{m-1}i_j2^j)$. 
%We denote by $\bar{x}=1-x$
%the binary complement of $x\in \{0,1\}$.
%For any given GBF $f$ in $m$ variables, we denote the function $f(1-x_0,1-x_1,\hdots,1-x_{m-1})$ or $f(\bar{x}_0,\bar{x}_1,\hdots,\bar{x}_{m-1})$  by $\tilde{f}$, 
%where $\bar{x}=1-x$ denotes the binary complement of $x\in \{0,1\}$.
%For a $\mathbb{Z}_q$-valued sequence $\textbf{e}$,
%$\bar{\textbf{e}}$ denotes the sequence $(q/2,q/2,\cdots,q/2)-\textbf{e}$.
%For a complex-valued sequence $\textbf{a}$, $\tilde{\textbf{a}}$ is obtained by reversing $\textbf{a}$ and $\textbf{a}^*$ is its complex conjugate.
\begin{definition}[Reed-Muller Code]
%A linear code over $\mathbb{Z}_q$ of length $L$ is closed
%under linear combinations of sequences (called codewords). Corresponding to any such code $\zeta$ there is a generator matrix
%$\text{G}$. Linear combinations of the rows of $\text{G}$ generate the code. For any fixed sequence $\textbf{a}$ of length $L$, $\textbf{a}+\zeta$
%denotes a coset of $\zeta$ and $\textbf{a}$ is said to be a coset representative of $\zeta$. 
A set of sequences which are obtained from 
the GBFs of $m$ variables $x_0, x_1, \hdots, x_{m-1}$ of order no greater than $r$ over $\mathbb{Z}_q$ is said to be $r$th order RM code over 
$\mathbb{Z}_q$ and is denoted by RM$_q(r,m)$.
RM$_q(r,m)$ is said to be the $r$th order RM code
%whose codewords are $\mathbb{Z}_q$-valued sequences identified with the GBFs of order
%at most $r$ of the variables $x_0, x_1, \hdots, x_{m-1}$ over $\mathbb{Z}_q$. 
%The rows of generator matrix $\text{G}$ for RM$_q(r,m)$ are $\mathbb{Z}_q$-valued sequences corresponding
%to distinct monomials of degree at most $r$ over the variables $x_0, x_1, \hdots, x_{m-1}$. The reader is referred to \cite{pater2000} for more details.
\end{definition}
%\begin{example}
%Consider \textnormal{RM}$_2(2,3)$, generated by vectors corresponding to the monomials of degree at most $2$ in variables
%$x_0$, $x_1$ and $x_2$. The generator matrix $\text{G}$ of \textnormal{RM}$_2(2,3)$ is given as follows.
%\begin{equation}\nonumber
%\begin{bmatrix}
% 11111111\\
% 01010101\\
% 00110011\\
% 00001111\\
% 00010001\\
% 00000101\\
% 00000011
%\end{bmatrix}
%\quad
%\begin{matrix}
% 1\\
% x_0\\
% x_1\\
% x_2\\
% x_0x_1\\
% x_0x_2\\
% x_1x_2
%\end{matrix}
%\end{equation}
%\end{example}
\subsection{Existing Constrcutions of CC, CCC, and ZCCS}
Some lemmas has been presented in this subsection and we also introduce some notations which will be used for our proposed constructions.
\begin{lemma}[\cite{pater2000}]\label{lemma1}
Let $f,g$ are two GBFs.
Assume
$\textbf{d}=(d_0, d_1,\hdots, d_{k-1})\in\{0,1\}^k$ and
$\{i_0,i_1,\cdots,i_{l-1}\}$ be a set of $l$ indices such that $0\leq i_0<i_1<\cdots< i_{l-1}<m$ and has no intersection with  $J$.
Let $\textbf{y}=(x_{i_0},x_{i_1},\hdots, x_{i_{l-1}})$, then
\begin{equation}
\begin{split}
\mathscr{C}&\big(\psi(f\arrowvert_{\textbf{x}_J=\textbf{c}}),\psi(g\arrowvert_{\textbf{x}_J=\textbf{d}})\big)(\tau)\\&=\displaystyle\sum_{\textbf{c}_1,\textbf{c}_2}
\mathscr{C}\big(\psi(f\arrowvert_{\textbf{x}_J\textbf{y}=\textbf{cc}_1}),\psi(g\arrowvert_{\textbf{x}_J\textbf{y}=\textbf{dc}_2})\big)(\tau).
\end{split}
\end{equation}
\end{lemma}
\vspace{0.2cm}
\begin{lemma}[{\cite[Th. 12]{pater2000}}]\label{lemmapat}
 Let $f$ is a second-order GBF and $G(f\arrowvert_{\textbf{x}_J=\textbf{c}})$ is a path with $x_\gamma$ as it's one of the end vertices for all
 $\textbf{c}\in\{0,1\}^k$. Assume all the edges in the path have the identical weights of $q/2$. 
 %Let say $G(f)$ contains a set of $k$ distinct
 %vertices labeled $j_0, j_1, \hdots, j_{k-1}$. It has the property that deleting the $k$ vertices and corresponding their edges results in a path. We further assume
 %that $x_\gamma$ is one of the end vertices of the path and all the edges of the path are of identical weight $q/2$. 
 Then for any choice of $g_i$, $g'\in $ $\mathbb{Z}_q$
\begin{equation}
\left\{f+\frac{q}{2}\left(\sum_{\alpha=0}^{k-1}u_{\alpha}x_{j_{\alpha}}+ux_\gamma    \right): u_{\alpha}, u \in \{0,1\} \right\}  
\end{equation}
is a CC of size $2^{k+1}$.
\end{lemma}
\vspace{0.2cm}
%\textit{Lemma} \ref{lemma1}
%\section{Proposed Constructions}
\begin{lemma}[\cite{arthina}]{(Construction of CCC)}\label{lemma3}\\
Let $f$ is a second-order GBF which has the same property as defined in \textit{Lemma} \ref{lemmapat}. 
%Suppose
%$G(f)$ contains a set of $k$ distinct vertices labeled $j_0,j_1,\hdots,j_{k-1}$ with the property
%that deleting those $k$  vertices and all their edges results in a path with $q/2$ being the weight of every edge of the path.
Consider $(t_0,t_1,\hdots, t_{k-1})$ is the binary representation of the integer $t$. Define the CC $C_t$ to be
\begin{equation}
 \displaystyle \left\{ f\!+\!
 \frac{q}{2}\left(\displaystyle{\sum_{\alpha=0}^{k-1}} u_{\alpha}x_{j_{\alpha}}\!+\!\displaystyle{\sum_{\alpha=0}^{k-1}} t_{\alpha}x_{j_{\alpha}}
 \!+\!ux_{\gamma}\right): u,u_\alpha\in\{0,1\}     \right\},
\end{equation}
and $\bar{C}_{2^k+t}$ to be
\begin{equation}
\displaystyle \left\{ \tilde{f}\!+\!
 \frac{q}{2}\left(\displaystyle{\sum_{\alpha=0}^{k-1}} u_{\alpha}\bar{x}_{j_{\alpha}}\!+\!\displaystyle{\sum_{\alpha=0}^{k-1}} t_{\alpha}\bar{x}_{j_{\alpha}}
 \!+\!\bar{u}x_{\gamma}\right): u,u_\alpha\in\{0,1\}     \right\}.
 \end{equation}
 %where $\gamma$ be the label of either end vertex in the path.
 Then
 \begin{equation}
  \left\{\psi(C_t):0\leq t<2^k\right\}\cup \left\{\psi^*(\bar{C}_{2^k+t}):0\leq t<2^k\right\}
 \end{equation}
 generate a set of CCC, where $\psi^*(\cdot)$ is the complex conjugate of $\psi(\cdot)$.
\end{lemma}
\vspace{0.2cm}
%\textit{Lemma} \ref{lemma3} will be used in \textit{Theorem} \ref{theorem2} to show that the construction of CCC in \cite{arthina} is a special case
%of our construction.
Before presenting the next lemmas, define $I_t=\{0,1,\hdots,t-1\}$. Therefore, $I_m$ is a set of indices of the variables 
$x_0,x_1,\hdots,x_{m-1}$. We assume 
%\begin{equation}\nonumber
% \begin{cases}
%  J:&\left\{j_0,j_1,\hdots,j_{k-1}\right\}\subset I_m;\\
  $J':\left\{j'_0,j'_1,\hdots,j'_{m-k-1}\right\}\subset I_m\setminus J$,
% \end{cases}
%\end{equation}
and
\begin{equation}
 \begin{cases}
  W:&\left\{w_0,w_1,\hdots,w_{k-1}\right\}\subset I_n, \{w_k\}\subset I_n\setminus W;\\
  W':&\left\{w'_0,w'_1,\hdots,w'_{n-k-2}\right\}\subset I_n\setminus \left\{W,w_k\right\}.
 \end{cases}
\end{equation}
The above defined sets will be used for representation of 
below binary vectors: $\textbf{x}=\left( x_0,x_1,\hdots,x_{m-1}\right)$, 
%$\textbf{x}_J=\left(x_{j_0},x_{j_1},\hdots,x_{j_{k-1}}\right)$,
$\textbf{u}=\left(u_0,u_1,\hdots,u_{n-1}\right)$, $\textbf{u}_W=\left(u_{w_0},u_{w_1},\hdots,u_{w_{k-1}}\right)$, and 
$\textbf{u}_{W'}=\left(u_{w'_0},u_{w'_1},\hdots,u_{w'_{n-k-2}}\right)$. 
\vspace{0.2cm}
\begin{lemma}[\cite{p_tcom}]{(Construction of ZCCS)}\label{lm1}\\
Let $f$ is a second-order GBF. Assume, $G(f\arrowvert_{\textbf{x}_J=\textbf{c}})$ 
%contains a set of vertices 
%$x_{j_0},x_{j_1},\hdots,x_{j_{k-1}}$ with the property that deleting those vertices and all their associated edges from $G(f)$, the resulting graph
consists of a path with $x_\gamma$ as one of its end vertices and $p$ isolated vertices $x_{m-p},x_{m-p+1},\hdots,x_{m-1}$. Let ${t_0,t_1,\hdots,t_{k+p-1}}$ be the binary vector 
representation of $t$ then the ordered set $S_t$ is defined as 
\begin{equation}
\begin{split}
 \displaystyle\left \{f+
 \frac{q}{2}\left(\displaystyle{\sum_{\alpha=0}^{k-1}} u_{\alpha}x_{j_{\alpha}}+\displaystyle{\sum_{\alpha=0}^{k-1}} t_{\alpha}x_{j_{\alpha}}+
 \displaystyle{\sum_{\alpha=0}^{p-1}} t_{k+\alpha}x_{m-p+\alpha}\right.\right.\\ \left.\left.+u_k x_{\gamma}\right): u_k,u_\alpha\in \mathbb{Z}_2     \right\},
\end{split}
\end{equation}
and the counterpart set $\bar{S}_t$ to be
\begin{equation}
\begin{split}
 \displaystyle\left \{\tilde{f}+
 \frac{q}{2}\left(\displaystyle{\sum_{\alpha=0}^{k-1}} u_{\alpha}\bar{x}_{j_{\alpha}}+\displaystyle{\sum_{\alpha=0}^{k-1}}
 t_{\alpha}\bar{x}_{j_{\alpha}}+ \displaystyle{\sum_{\alpha=0}^{p-1}} t_{k+\alpha}\bar{x}_{m-p+\alpha}\right.\right.\\ \left. \left.+\bar{u_k}x_{\gamma}\right): u_k,u_\alpha\in\{0,1\}\right\}.\qquad\qquad\qquad
\end{split}
\end{equation}
Then
\begin{equation}\nonumber
 \left\{\psi({S}_t): 0\leq t\leq 2^{k+p}-1\right\}\cup \left\{\psi^*(\bar{S}_t): 0\leq t\leq 2^{k+p}-1\right\},
 \end{equation}
 form $\left(2^{k+p+1},2^{m-p}\right)$-$\text{ZCCS}_{2^{k+1}}^{2^m}$.
\end{lemma}
\begin{lemma}[\cite{liu2011}]\label{lemma5}
 For any $(K,Z)$-$\text{ZCCS}_M^L$, the theoretical bound is given by
\begin{equation}\label{op}
  K\leq M\lfloor L/Z\rfloor,
  \end{equation}
 % where $Z$ is the ZCZ width, $K$ is the number of $Z$-complementary codes, $M$ is the number of constituent sequences in a $Z$-complementary code and
 % $L$ is the length of each constituent sequence.
We call $(K,Z)$-$\text{ZCCS}_M^L$ is optimal if $K= M\lfloor L/Z\rfloor$.
\end{lemma}
\section{Proposed New Construction of ZCCS With Maximum Column Sequence PMEPR $2$}
Before presenting our propose construction, we first define the following notations:
\begin{equation}\nonumber
\begin{split}
 \begin{cases}
 \mathcal{Z}:&\left\{m-p,m-p+1,\hdots,m-1\right\}\subset I_m,\\ &\textnormal{where}~0\leq p\leq m-k-1\\
  J_1:&\left\{j_0,j_1,\hdots,j_{k-1}\right\}\subset I_m\setminus \mathcal{Z}\\
  J'_1:&\left\{j'_0,j'_1,\hdots,j'_{m-k-p-1}\right\}\subset I_m\setminus J_1\cup \mathcal{Z}.
 \end{cases}
 \end{split}
\end{equation}
Here, $J_1$ is equals to $J$, when $\mathcal{Z}=\phi$.
We also define the following binary vectors:
$\textbf{x}_\mathcal{Z}=\left(x_{m-p},x_{m-p+1},\hdots,x_{m-1}\right)$ and $\textbf{x}_{J_1}=\left(x_{j_0},x_{j_1},\hdots,x_{j_{k-1}}\right)$.
Now, we present a lemma which will be used in our propose construction.
\begin{lemma}\label{lemmaz}
Let $f$ and $f'$ are two GBFs and $f'\arrowvert_{\textbf{x}_{J_1}=\textbf{c}}$ are given by
%of $m$ variables $x_0,x_1,\hdots,x_{m-1}$
%$(m\geq 2)$, such that for some $k$ $(0\leq k\leq m-p-2$, $p\geq 0)$,  $f\arrowvert _{\textbf{x}_{J_1}=\textbf{c}}$
%and $f'\arrowvert_{\textbf{x}_{J_1}=\textbf{c}}$ are given by
\begin{equation}\nonumber
\begin{split}
f\arrowvert _{\textbf{x}_{J_1}=\textbf{c}}&=P+L+g_{m-p}x_{m-p}+g_{m-p+1}x_{m-p+1}\\&~~~~~~~~~~~~~~~~~~~~~~+\cdots+g_{m-1}x_{m-1}+g',\qquad\qquad \\
f'\arrowvert _{\textbf{x}_{J_1}=\textbf{c}}&=f\arrowvert _{\textbf{x}_{J_1}=\textbf{c}}+\frac{q}{2}x_{\gamma_\textbf{c}},
\end{split}
\end{equation}
where
\begin{equation}\nonumber
\begin{split}
P&=\frac{q}{2}\displaystyle\sum_{\alpha=0}^{m-k-p-2}x_{j'_{\alpha}}x_{j'_{\alpha+1}},\\
L&=
 \displaystyle\sum_{\alpha=0}^{m-k-p-1}g_{j'_\alpha}x_{j'_{\alpha}},
 \end{split}
\end{equation}$g',g_{j'_\alpha}\in \mathbb{Z}_q$, $\alpha =0,1,\hdots,m-k-p-1$ and $\gamma_{\textbf{c}}$ is one of the end vertices of the
%the label of  either
path $G(P)$.
Then for fixed $\textbf{c}\in \{0,1\}^k$ and $\textbf{d}'\neq\textbf{d}''$, where $\textbf{d}'=(d'_1,d'_2,\hdots,d'_p)$ $\left(\in\{0,1\}^p\right)$ and 
$\textbf{d}''=(d''_1,d''_2,\hdots,d''_p)$ $\left(\{0,1\}^p\right)$, we have \\
%with $\psi(f)$ and $\psi(f')$, the vectors corresponding to $f$ and $f'$ respectively, and
 $\mathscr{C}\left(f\arrowvert _{\textbf{x}\textbf{x}'=\textbf{c}\textbf{d}'},f\arrowvert _{\textbf{x}\textbf{x}'=\textbf{c}\textbf{d}''}\right)(\tau)+\mathscr{C}\left(f'\arrowvert _{\textbf{x}\textbf{x}'=\textbf{c}\textbf{d}'},f'\arrowvert _{\textbf{x}\textbf{x}'=\textbf{c}\textbf{d}''}\right)(\tau)$
 \begin{equation}
 \begin{split}
  =\begin{cases}
\omega^{(d_1'-d_1'')g_{m-p}}+\cdots+(d_p'-d_p'')g_{m-1}2^{m-(k+p)+1},  \\~~~~~~~~~~~~~~\tau=(d_1'-d_1'')2^{m-p}+\cdots+(d_p'-d_p'')2^{m-1},\\
0, ~~~~~~~~~~~~\textnormal{otherwise}.
 \end{cases}
 \end{split}
 \end{equation}
 \end{lemma}
\begin{theorem}\label{thm1}
Let $f(\textbf{x},\textbf{u}_{W'}):\{0,1\}^{n+m-k-1}\rightarrow \mathbb{Z}_q$ and $h(\textbf{u}):\{0,1\}^n\rightarrow \mathbb{Z}_q$ are two 
GBFs of of degree greater than $1$. Suppose, $f$ has the property that 
$G\left(f\arrowvert_{\textbf{x}_{J_1}=\textbf{c},\textbf{u}_{W'}=\textbf{e}}\right)$, where $\textbf{c}\in \{0,1\}^k$ and $\textbf{e}\in \{0,1\}^{n-k-1}$, 
contains a Hamiltonian path whose vertices are specified by $J'_1$ and $\gamma_\textbf{c}$ is one of the end vertices. $p$ is the number of isolated vertices which are 
specified by $\mathcal{Z}$ and the edges in the path are having identical weights $q/2$. Also, let $(t_0,t_1,\hdots,t_{n+p-2})$ be binary representation of the integer $t$.   Define, the code $S_t$ to be 
\begin{equation}\label{mainsett1}
\begin{split}
 \left\{g+\frac{q}{2}\left(\displaystyle\sum_{\alpha=0}^{k-1}u_{w_\alpha}x_{j_\alpha}+u_{w_k}x_{\gamma_{\textbf{c}}}\right)
 \!\!+\!\!\frac{q}{2}\left(\sum_{\alpha=0}^{k-1}t_\alpha x_{j_\alpha}\right.\right.\\ \left.\left.+\sum_{\alpha=0}^{n-k-2}t_{\alpha+k}u_{w'_\alpha}\!\!+\!\!\sum_{\alpha=0}^{p-1}t_{\alpha+n-1}x_{m-p+\alpha}\right)\right\},
 \end{split}
\end{equation}
and the counterpart code $\bar{S}_t$ is defined as
\begin{equation}\label{mainsett2}
 \begin{split}
  \left\{\tilde{g}+\frac{q}{2}\left(\displaystyle\sum_{\alpha=0}^{k-1}u_{w_\alpha}\bar{x}_{j_\alpha}+\bar{u}_{w_k}x_{\gamma_{\textbf{c}}}\right)
 +\frac{q}{2}\left(\sum_{\alpha=0}^{k-1}t_\alpha \bar{x}_{j_\alpha}\right.\right.\\ \left.\left.+\sum_{\alpha=0}^{n-k-2}t_{\alpha+k}u_{w'_\alpha}+\sum_{\alpha=0}^{p-1}t_{\alpha+n-1}\bar{x}_{m-p+\alpha}\right)\right\}.
 \end{split}
\end{equation}
Then 
\begin{equation}\label{mainset}
 \{\psi({S}_t): 0\leq t\leq 2^{n+p-1}-1\}\cup \{\psi^*(\bar{S}_t): 0\leq t\leq 2^{n+p-1}-1\},
 \end{equation}
 form $(2^{n+p},2^{m-p})$-$\text{ZCCS}_{2^{n}}^{2^m}$ if
 \begin{equation}\label{cond}
  h\arrowvert_{\textbf{u}_{W'},\textbf{u}_Wu_{w_k}=\textbf{e},\textbf{b}b}\in \{\delta,\frac{q}{2}+\delta\},
 \end{equation}
where $\delta\in \mathbb{Z}_q$ and $b\textbf{b}\in \{0,1\}^{k+1}$.
\end{theorem}
\begin{IEEEproof}
Please see Appendix A.
\end{IEEEproof}
\vspace{0.2cm}
\begin{remark}[Construction of GBFs as Defined in \textit{Theorem} \ref{thm1}]\label{re3}
The GBF corresponding to \textit{Theorem} \ref{thm1} is given by $g(\textbf{x},\textbf{u})=f(\textbf{x},\textbf{u}_{W'})+h(\textbf{u})$, where
the function $f(\textbf{x},\textbf{u}_{W'})$ can be expressed as 
\begin{equation}\label{expgbf1}
 \begin{split}
 f&\left(\textbf{x},\textbf{u}_{W'}\right)=\\ &\!\!\!\!\!\!\frac{q}{2}\displaystyle\sum_{\textbf{c}\in \{0,1\}^k}\!\!\!\!\!\!\!\!\sum_{\alpha=0}^{m-k-p-2}\!\!\!\!x_{j'_{\pi_\textbf{c}(\alpha)}}x_{j'_{\pi_\textbf{c}(\alpha+1)}}
 \!\!\prod_{\alpha=0}^{k-1}x_{j_\alpha}^{c_\alpha}(1-x_{j_\alpha})^{1-c_\alpha}\\
 +&\sum_{r=1}^k\sum_{0\leq \alpha_0<\alpha_1<\cdots<\alpha_{r}< k}
  \!\!\!\!\!\!\!\!\!\!\!\!\!\!\!\!\!\!
\varrho_{j_{\alpha_0},j_{\alpha_1},\hdots,j_{\alpha_r}}x_{j_{\alpha_0}}x_{j_{\alpha_1}}\!\!\!\!\!\cdots x_{j_{\alpha_r}}\\
+&\sum_{r=1}^k\sum_{\alpha=0}^{m-k-p-1}\sum_{0\leq \alpha_0<\alpha_1<\cdots<\alpha_{r}< k}
  \!\!\!\!\!\!\!\!\!\!\!\!\!\!\!\!\!\!
\kappa_{j_{\alpha_0},j_{\alpha_1},\hdots,j_{\alpha_r}}^{\alpha}x_{j_{\alpha_0}}x_{j_{\alpha_1}}\!\!\!\!\!\cdots x_{j_{\alpha_r}}x_{j'_\alpha}\\
 +&\sum_{r=1}^k\sum_{\alpha=0}^{p-1}\sum_{0\leq \alpha_0<\alpha_1<\cdots<\alpha_{r}< k}
  \!\!\!\!\!\!\!\!\!\!\!\!\!\!\!\!\!\!
\varrho^{m-p+\alpha}_{j_{\alpha_0},j_{\alpha_1},\hdots,j_{\alpha_r}}x_{j_{\alpha_0}}x_{j_{\alpha_1}} \!\!\!\!\!\cdots x_{j_{\alpha_r}}x_{m-p+\alpha}\\
+&\sum_{r=1}^k\sum_{\alpha=0}^{n-k-2}\sum_{0\leq \alpha_0<\alpha_1<\cdots<\alpha_{r}< k}
  \!\!\!\!\!\!\!\!\!\!\!\!\!\!\!\!\!\!
\varrho'^{\alpha}_{j_{\alpha_0},j_{\alpha_1},\hdots,j_{\alpha_r}}x_{j_{\alpha_0}}x_{j_{\alpha_1}}\!\!\!\!\!\cdots x_{j_{\alpha_r}}x_{u_{w'_\alpha}}\\
+&\sum_{\alpha=0}^{m-1}x_\alpha\sum_{\beta=0}^{n-k-2}\vartheta_\alpha^\beta x_{u_{w'_\beta}}+\sum_{\alpha=0}^{m-1}g_\alpha x_\alpha+g',
 \end{split}
\end{equation}
where 
\begin{itemize}
\item $\pi_\textbf{c}$ ($\textbf{c}\in \{0,1\}^k$) is a permutation on the set $\{0,1,\hdots,m-k-p-1\}$, 
\item $\varrho_{j_{\alpha_0},j_{\alpha_1},\hdots,j_{\alpha_r}}\in \mathbb{Z}_q$ 
for $r=1,2,\hdots,k,~0\leq \alpha_0<\alpha_1<\cdots<\alpha_{r}< k$, 
\item $\kappa_{j_{\alpha_0},j_{\alpha_1},\hdots,j_{\alpha_r}}^{\alpha}\in\mathbb{Z}_q$, for $r=1,2,\hdots,k,~0\leq \alpha_0<\alpha_1<\cdots<\alpha_{r}< k
,\alpha=0,1,\hdots,m-k-p-1$,
\item $\varrho^{m-p+\alpha}_{j_{\alpha_0},j_{\alpha_1},\hdots,j_{\alpha_r}}\in\mathbb{Z}_q$ for $r=1,2,\hdots,k,~0\leq \alpha_0<\alpha_1<\cdots<\alpha_{r}< k,~
\alpha=0,1,\hdots,p-1$, 
\item $\varrho'^{\alpha}_{j_{\alpha_0},j_{\alpha_1},\hdots,j_{\alpha_r}}\in \mathbb{Z}_q$ for 
$r=1,2,\hdots,k,~0\leq \alpha_0<\alpha_1<\cdots<\alpha_{r}< k$, 
\item $\vartheta_\alpha^\beta\in \mathbb{Z}_q,~\alpha=0,1,\hdots,m-1,~\beta=
0,1,\hdots,n-k-2$, $g_\alpha\in\mathbb{Z}_q$, $g\in\mathbb{Z}_q$, 
\end{itemize}
and the term
\begin{equation}\label{teq1}
 \begin{split}
  \sum_{r=1}^k\sum_{\alpha=0}^{p-1}\sum_{0\leq \alpha_0<\alpha_1<\cdots<\alpha_{r}< k}
  \!\!\!\!\!\!\!\!\!\!\!\!\!\!\!\!\!\!
\varrho^{m-p+\alpha}_{j_0,j_1,\hdots,j_r}x_{j_{\alpha_0}}x_{j_{\alpha_1}}\!\!\!\!\!\cdots x_{j_{\alpha_r}}x_{m-p+\alpha},
 \end{split}
\end{equation}
is denoted by $\mathcal{G}_{\textbf{x}_{J_1}\textbf{x}_\mathcal{Z}}$ in the proof of \textit{Theorem} \ref{thm1} and also will be used in the construction of IGC code set.
The function $h(\textbf{u})$ can be taken to be of any order. For our desired result (as we are interested to design ZCCS of maximum 
column sequence PMEPR $2$) we take the function as follows:
%The function $h(\textbf{u})$ can be expressed as
\begin{equation}\label{expgbf2}
 h(\textbf{u})=\frac{q}{2}\displaystyle\sum_{\alpha=0}^{n-2}u_{\pi(\alpha)}u_{\pi(\alpha+1)}+\sum_{\alpha=0}^{n-1}\lambda_\alpha u_\alpha+\lambda',
\end{equation}
where $\pi$ is a permutation on the set $\{0,1,\hdots,n-1\}$, $\lambda_\alpha\in \mathbb{Z}_q$ for $\alpha=0,1,\hdots,n-1$, and 
$\lambda'\in\mathbb{Z}_q$.

It is noted that we design $f$ such a way that $G(f\arrowvert_{\textbf{x}=\textbf{c}})$ contains path over the vertices 
$x_{j'_{\pi_\textbf{c}(0)}},x_{j'_{\pi_\textbf{c}(1)}},\hdots,x_{j'_{\pi_\textbf{c}(m-k-p-1)}}$ and $p$ isolated
vertices $x_{m-p},x_{m-p+1},\hdots,x_{m-1}$. It is also noted that $x_{j'_{\pi_\textbf{c}(0)}}$ and $x_{j'_{\pi_\textbf{c}(m-k-p-1)}}$ are the 
end vertices in the path which is contained in $G(f\arrowvert_{\textbf{x}=\textbf{c}})$ and in \textit{Theorem} \ref{thm1}, 
$\gamma_{\textbf{c}}$ is taken as either $j'_{\pi_\textbf{c}(0)}$ or $j'_{\pi_\textbf{c}(m-k-p-1)}$.
\end{remark}
\vspace{0.2cm}
\begin{corollary}\label{re1}
From \textit{Theorem} \ref{thm1}, we obtain $K$ ($=2^{n+p}$) codes where each code contains $M$ ($=2^n$) constituent sequences of length
$L$ ($=2^m$) with ZCZ width $Z$ ($=2^{m-p}$). The code set also satisfies the equality $K=\frac{LM}{Z}$ and thus, $(2^{n+p},2^{m-p})$-$\text{ZCCS}_{2^{n}}^{2^m}$
is an optimal ZCCS.
%with respect to the theoretical bound given in \textit{Lemma} \ref{lemma5}.
\end{corollary}
\begin{corollary}\label{re2}
 In \textit{Theorem} \ref{thm1}, we take $f(\textbf{x},\textbf{u}_{W'})$ and $h(\textbf{u}$ as given in
 (\ref{expgbf1}) and (\ref{expgbf2}) respectively. As, $g(\textbf{x},\textbf{u})=f(\textbf{x},\textbf{u}_{W'})+h(\textbf{u}$ and $G\left(g(\textbf{c},\textbf{u})\right)$ is a path 
 over the vertices $u_0,u_1,\hdots,u_{n-1}$ for all $\textbf{c}\in \{0,1\}^m$, the maximum column sequence PMEPR of
 the $(2^{n+p},2^{m-p})$-$\text{ZCCS}_{2^{n}}^{2^m}$ is $2$.
 \begin{IEEEproof}
  We recall the set $S_t$ given in (\ref{mainsett1}) and assume
  \begin{equation}\label{graphf}
   \begin{split}
    \mathcal{F}^t_{\textbf{x},\textbf{u}}&=g+\frac{q}{2}\left(\displaystyle\sum_{\alpha=0}^{k-1}u_{w_\alpha}x_{j_\alpha}+u_{w_k}x_\gamma\right)\\& \qquad
 +\frac{q}{2}\left(\sum_{\alpha=0}^{k-1}t_\alpha x_{j_\alpha}+\sum_{\alpha=0}^{n-k-2}t_{\alpha+k}u_{w'_\alpha}\right.\\ &\qquad\qquad\left.+\sum_{\alpha=0}^{p-1}t_{\alpha+n-1}x_{m-p+\alpha}\right)
   \end{split}
  \end{equation}
Let $\textbf{x}_0,\textbf{x}_1,\hdots,\textbf{x}_{m-1}$ are the binary vector representations of $0,1,\hdots,2^m-1$, and 
$\textbf{u}_0,\textbf{u}_1,\hdots,\textbf{u}_{n-1}$ are the binary vector representations of $0,1,\hdots,2^n-1$.
Therefore, 
%\begin{equation}
%\psi(S_t)=\begin{bmatrix}
%           \omega^{\mathcal{F}^t_{\textbf{x}_0,\textbf{u}_0}}~~\omega^{\mathcal{F}^t_{\textbf{x}_1,\textbf{u}_0}}~\cdots~\omega^{\mathcal{F}^t_{\textbf{x}_{2^m-1},\textbf{u}_0}}\\
%           \omega^{\mathcal{F}^t_{\textbf{x}_0,\textbf{u}_1}}~~\omega^{\mathcal{F}^t_{\textbf{x}_1,\textbf{u}_1}}~\cdots~\omega^{\mathcal{F}^t_{\textbf{x}_{2^m-1},\textbf{u}_1}}\\
%           \vdots~~~~~~~~~\vdots~~~~~~~~~~\vdots\\
%           \omega^{\mathcal{F}^t_{\textbf{x}_0,\textbf{u}_{2^n-1}}}~~\omega^{\mathcal{F}^t_{\textbf{x}_1,\textbf{u}_{2^n-1}}}~\cdots~\omega^{\mathcal{F}^t_{\textbf{x}_{2^m-1},\textbf{u}_{2^n-1}}}\\
%          \end{bmatrix}
%\end{equation}
\begin{equation}
\psi(S_t)=\begin{bmatrix}
           \omega^{\mathcal{F}^t_{\textbf{x}_0,\textbf{u}_0}}&\omega^{\mathcal{F}^t_{\textbf{x}_1,\textbf{u}_0}}~~\cdots&\omega^{\mathcal{F}^t_{\textbf{x}_{2^m-1},\textbf{u}_0}}\\
           \omega^{\mathcal{F}^t_{\textbf{x}_0,\textbf{u}_1}}&\omega^{\mathcal{F}^t_{\textbf{x}_1,\textbf{u}_1}}~~\cdots&\omega^{\mathcal{F}^t_{\textbf{x}_{2^m-1},\textbf{u}_1}}\\
           \vdots&~~~\vdots~~~~~~~~~~~~\ddots&\vdots\\
           \omega^{\mathcal{F}^t_{\textbf{x}_0,\textbf{u}_{2^n-1}}}&\omega^{\mathcal{F}^t_{\textbf{x}_1,\textbf{u}_{2^n-1}}}~~\cdots&\omega^{\mathcal{F}^t_{\textbf{x}_{2^m-1},\textbf{u}_{2^n-1}}}\\
          \end{bmatrix}
\end{equation}
The $j$th column of the code $\psi(S_t)$ is denoted by $\psi(\mathcal{F}^t_{\textbf{x}_j,\textbf{u}})$ and given by $\left[\omega^{\mathcal{F}^t_{\textbf{x}_j,\textbf{u}_0}}~\omega^{\mathcal{F}^t_{\textbf{x}_j,\textbf{u}_1}}~\cdots~\omega^{\mathcal{F}^t_{\textbf{x}_j,\textbf{u}_{2^n-1}}}\right]^T$,
where $j=0,1,\hdots,2^{m}-1$. If we setup $k=0$ in \textit{Lemma} \ref{lemmapat}, i.e., if $G(f)$ is a path, the set $\left\{\psi(f),\psi(f+\frac{q}{2}x_\gamma)
\right\}$ forms a GCP. From (\ref{graphf}), it is clear that $G(\mathcal{F}^t_{\textbf{x}_j,\textbf{u}})$ is the same as $G(g(\textbf{x}_j,\textbf{u}))$
which is a path for all $j$ over the vertices $u_0,u_1,\hdots,u_{n-1}$. Therefore, by using \textit{Lemma} \ref{lemmapat}, the set
$\left\{\psi(\mathcal{F}^t_{\textbf{x}_j,\textbf{u}}), \psi(\mathcal{F}^t_{\textbf{x}_j,\textbf{u}}+u_\gamma)\right\}$ forms a GCP where
$u_\gamma$ is assumed to be one of the end vertices of the path $G(\mathcal{F}^t_{\textbf{x}_j,\textbf{u}})$. Therefore, each column of the code 
$\psi(S_t)$ lies in a GCP and hence the maximum PMEPR is $2$. Similarly, we can show that the maximum PMEPR of each column of the code
$\psi^*(\bar{S}_t)$ is $2$. Therefore, the maximum column sequence PMEPR of the ZCCS $\{\psi({S}_t): 0\leq t\leq 2^{n+p-1}-1\}\cup \{\psi^*(\bar{S}_t): 0\leq t\leq 2^{n+p-1}-1\}$
or, $(2^{n+p},2^{m-p})$-$\text{ZCCS}_{2^{n}}^{2^m}$ is $2$.
\end{IEEEproof}
 %is a GDJ sequence. Therefore. the maximum column sequence PMEPR of  
\end{corollary}
\begin{remark}\label{spl1}
 For the case, $\mathcal{Z}=\phi$, the result of \textit{Theorem} \ref{thm1} reduces to the result given in \cite{uda2014}.
 %which appears
 %as \textit{Lemma} \ref{le6} in this paper. 
 Therefore, the
 construction given in \cite{uda2014} appears as special case of the proposed ZCCS construction.
\end{remark}
\begin{remark}
 For $W\cup W'\cup \{w_k\}=\phi$, the function $g$ given in \textit{Theorem} \ref{thm1} reduces to $g=f$. If we consider the degree of $f$
 is $2$ and $W\cup W'\cup \{w_k\}=\phi$, the result of \textit{Theorem} \ref{thm1} reduces to the result given in \cite[Th. 2]{p_tcom}. 
 Therefore, the construction given in \cite{p_tcom} and \cite{pater2000}, which appear as \textit{Lemma} \ref{lemmapat} and \textit{Lemma} \ref{lm1} repectively in this paper, are special cases of our proposed construction. 
 %It is observed that the constrcution
 %given in \cite{pater2000} and \cite{Davis1999} are special cases of the construction given in \cite{p_tcom}. Hence, the constrcutions given
 %in \cite{pater2000} and \cite{Davis1999} are also special cases of our proposed construction.
\end{remark}
Below, we present an example to illustrate \textit{Theorem} \ref{thm1}.
\begin{example}
We consider $m=5,n=2, k=1,p=1, W=\{w_0\}=\{0\}$, and $\{w_1\}=\{1\}$. Therefore, $W'$ is a null set. We also consider 
Let $f:\{0,1\}^5\rightarrow \mathbb{Z}_4$ be a GBF given by
\begin{equation}
\begin{split}
f=&2((1-x_0)(x_3x_2+x_2x_1)+x_0(x_1x_3+x_3x_2)+x_0x_4\\
&+x_1+2x_2+2x_3+x_4+2,
\end{split}
\end{equation}
and the function $h(u_0,u_1)$ is given by
\begin{equation}
 h(u_0,u_1)=2u_0u_1.
\end{equation}
From $f$, it is clear that 
$J_1=\{0\},J_1'=\{1,2,3\},$ and $\mathcal{Z}=\{4\}$. Therefore, from \textit{Theorem} \ref{thm1} and \textit{Corollary} \ref{re2}, we obtain 
$(8,16)$-$\text{ZCCS}_{4}^{32}$ with maximum column sequence PMEPR $2$. The $(8,16)$-$\text{ZCCS}_{4}^{32}$ is given in TABLE I. In TABLE I,
$\psi(S_t)$ and $\psi^*(\bar{S}_t)$ are obtained by following (\ref{mainsett1}) and (\ref{mainsett2}).
\begin{table*}[!htbp]
\centering
%\tiny
\begin{tabular}{ |c|c| }
 \hline
 \multicolumn{2}{|c|}{$(8,16)$-$\text{ZCCS}_{4}^{32}$} \\
 \hline
 $\psi(S_0)$&$\psi(S_1)$ \\
 \hline
 $2   2  3   3   0   0      3    1   0   0    1  3      0   0   3   3   3    1      0    2    1   3  0   0    1   3    2    2    1   3   0    2$&$2  0   3    1  0   2     3   3  0   2   1  1     0   2  3    1  3      3   0   0   1   1   0       2    1   1    2   0   1        1   0   0$\\
 $2   0  3   1   0   2      3    3   0   2    1  1      0   2   3   1   3    3      0    0    1   1  0   2    1   1    2    0    1   1   0    0$&$2  2   3    3  0   0     3   1  0   0   1  3     0   0  3    3  3      1   0   2   1   3    0       0    1   3    2   2   1        3   0   2$\\
 $2   2  1   1   0   0      1    3   0   0    3  1      0   0   1   1   3    1      2    0    1   3  2   2    1   3    0    0    1   3   2    0$&$2  0   1    3  0   2     1   1  0   2   3   3     0   2  1    3  3      3   2   2   1   1  2       0    1   1    0   2   1        1   2   2$\\
 $0   2  3   1   2   0      3    3   2   0    1  1      2   0   3   1   1    1      0    0    3   3  0   2    3   3    2    0    3   3   0    0$&$0  0   3    3  2   2     3   1  2   2   1   3     2   2  3    3  1      3   0   2   3   1  0       0    3   1    2   2   3        1   0   2$\\
 \hline
 $\psi(S_2)$&$\psi(S_3)$\\
 \hline
 $2   2   3  3   0  0   3    1   0   0   1  3   0   0   3  3    1   3    2   0  3    1  2   2   3    1   0   0   3    1   2   0$ & $2   0   3   1  0    2   3  3   0   2   1   1   0   2   3   1    1   1   2   2   3   3   2   0   3   3   0   2   3   3   2   2$ \\
 $2   0   3  1   0  2   3    3   0   2   1   1   0   2   3  1    1   1    2   2  3    3  2   0   3    3   0   2   3    3   2   2$ & $2   2   3   3  0    0   3  1   0   0   1  3   0   0   3   3    1   3   2   0   3   1  2   2   3   1   0   0   3   1   2   0$ \\
 $2   2   1  1   0  0   1    3   0   0   3   1   0   0   1  1    1   3    0   2  3    1  0   0   3    1   2   2   3    1   0   2$ & $2   0   1   3  0    2   1  1   0   2   3   3   0   2   1   3    1   1   0   0   3   3  0   2   3   3   2   0   3   3   0   0$  \\
 $0   2   3  1   2  0   3    3   2   0   1  1   2   0   3  1    3   3    2   2  1    1  2   0   1    1   0   2   1    1   2   2$ & $0   0   3   3  2    2   3  1   2   2   1  3   2   2   3   3    3   1   2   0   1   3   2   2   1   3   0   0   1   3   2   0$ \\
  \hline
  $\psi^*(\bar{S}_0)$&$\psi^*(\bar{S}_1)$\\
  \hline
  $2   0   3    1   2   2   3   1   0   0   3  1   2   0   1   3    1    1   2   2   1  3  2  2   3    1  2   2   1   1   0   0$ & $0   0   1    1   0   2   1   1  2    0   1   1    0   0   3   3   3   1   0   2  3   3   0   2   1    1   0   2   3    1   2   0$ \\
  $ 0   0   1    1   0   2   1   1   2   0   1  1   0   0   3   3    3    1   0   2   3  3   0  2   1    1  0   2   3   1   2   0$ & $2   0   3    1   2   2   3   1  0    0   3  1    2   0   1   3   1   1   2   2  1   3  2   2   3    1   2   2   1    1   0   0$\\
  $2   0   1    3   2   2   1   3   0   0   1  3   2   0   3   1    1    1   0   0   1  3  0  0   3    1  0   0   1   1   2   2$ & $0   0   3    3   0   2   3   3  2    0   3  3    0   0   1   1   3   1   2   0  3   3  2   0   1    1   2   0   3    1   0   2$\\
  $2   2   1    1   2   0   1   1   0   2   1  1   2   2   3   3    1    3   0   2   1  1   0  2   3    3  0   2   1   3   2   0$ & $0   2   3    1   0   0   3   1  2    2   3   1    0   2   1   3   3   3   2   2  3   1   2   2   1    3   2   2   3    3   0   0$ \\
  \hline
  $\psi^*(\bar{S}_3)$&$\psi^*(\bar{S}_4)$\\
  \hline
  $0   2   1   3   0   0   1   3   2   2   1  3   0   2   3   1   1   1   2   2   1   3  2   2   3   1   2   2   1   1   0   0$ & $2   2   3   3   2   0   3   3   0   2   3  3   2   2   1   1   3   1   0   2   3   3   0   2    1   1   0   2   3   1   2   0$ \\
  $2   2   3   3   2   0   3   3   0   2   3  3   2   2   1   1   3   1   0   2   3   3  0   2   1   1   0   2   3   1   2   0$ & $0   2   1   3   0   0   1   3   2   2   1   3   0   2   3   1   1   1   2   2   1   3   2   2    3   1   2   2   1   1   0   0$ \\
  $ 0   2   3   1   0   0   3   1   2   2   3  1   0   2   1   3   1   1   0   0   1   3   0   0   3   1   0   0   1   1   2   2$ & $2   2   1   1   2   0   1   1   0   2   1  1   2   2   3   3   3   1   2   0   3   3  2   0    1   1   2   0   3   1   0   2$ \\
  $ 0   0   3   3   0   2   3   3   2   0   3  3   0   0   1   1   1   3   0   2   1   1   0   2   3   3   0   2   1   3   2   0$ & $2   0   1   3   2   2   1   3   0   0   1  3   2   0   3   1   3   3   2   2   3   1  2   2    1   3   2   2   3   3   0   0$ \\
  \hline
\end{tabular}
\caption{Optimal ZCCS over the alphabet $\mathbb{Z}_4$ with maximum column sequence PMEPR $2$.}
\end{table*}
\begin{figure}\label{gig1}
\centering
\includegraphics[height=6cm,width=8cm]{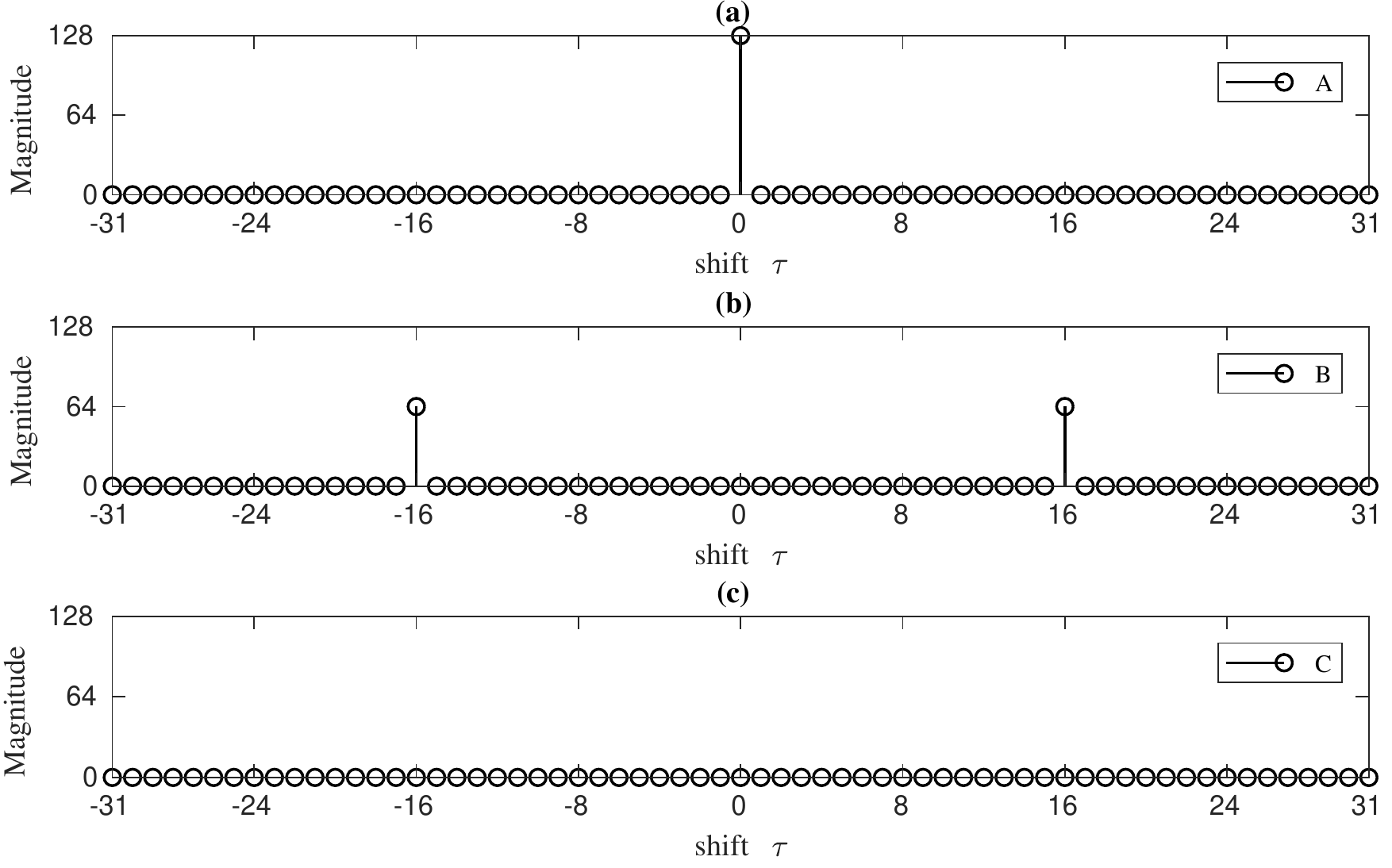}
\caption{Correlation plots of $(8,16)$-$\text{ZCCS}_{4}^{32}$ in TABLE I.}
%\caption{The graph of the quadratic form $x_2x_3+x_3x_1+x_0x_2+x_0x_3+x_0x_1+x_0x_4$.}
\end{figure}
In Fig. 1, Fig. 1-(a) represents AACF of any codes given in TABLE I, and Fig. 1-(b) and Fig. 1-(c) represent the ACCFs between two distinct
codes in TABLE I.
\end{example}
\section{Proposed New Construction of IGC Code Set With Maximum Column Sequence PMEPR $2$}
In this section, a new construction of IGC code set has been presented by using GBFs of order no less than $2$. We recall 
$(t_0,t_1,\hdots,t_{n+p-2})$ be binary representation of the integer $t$ ($0\leq t\leq 2^{n+p-1}-1$), where 
$\textbf{t}_1=(t_0,t_1,\hdots,t_{k-1}), \textbf{t}_2=(t_k,t_{k+1},\hdots,t_{n-2})$, and $\textbf{t}_3=(t_{n-1},t_n,\hdots,t_{n+p-2})$. We 
assume $(t_0,t_1,\hdots, t_{n-2})$, or, $\textbf{t}_1\textbf{t}_2$ be the binary vector representation of $\mathcal{T}$ 
($0\leq \mathcal{T}\leq 2^{n-1}-1$). Also $(t_0,t_1,\hdots,t_{n+p-2})$ and $\textbf{t}_1\textbf{t}_2\textbf{t}_3$ 
represent the same binary vectors. Now, we define the following code groups $\mathcal{I}_0,\mathcal{I}_1,\hdots,\mathcal{I}_{2^{n-1}-1},
\bar{\mathcal{I}}^*_{0},\bar{\mathcal{I}}^*_{1},\hdots,\bar{\mathcal{I}}^*_{2^{n-1}-1}$ as follows:
\begin{equation}\label{igc1}
 \begin{split}
  \mathcal{I}_\mathcal{T}=\left\{\psi(S_t):\textbf{t}_3\in\{0,1\}^p\right\},
 \end{split}
\end{equation}
and 
\begin{equation}\label{igc2}
 \begin{split}
  \bar{\mathcal{I}}^*_\mathcal{T}=\left\{\psi^*(\bar{S}_t):\textbf{t}_3\in\{0,1\}^p\right\},
 \end{split}
\end{equation}
where $0\leq\mathcal{T}\leq 2^{n-1}-1$. 
\begin{theorem}\label{thm2}
Let $f(\textbf{x},\textbf{u}_{W'}):\{0,1\}^{n+m-k-1}\rightarrow \mathbb{Z}_q$ and $h(\textbf{u}):\{0,1\}^n\rightarrow \mathbb{Z}_q$ be two $q$-ary
GBFs as defined in (\ref{expgbf1}) and (\ref{expgbf2}) respectively. We also assume $\mathcal{G}_{\textbf{x}_{J_1}\textbf{x}_\mathcal{Z}}=0$. Then
the code groups $\mathcal{I}_0,\mathcal{I}_1,\hdots,\mathcal{I}_{2^{n-1}-1},
\bar{\mathcal{I}}^*_{0},\bar{\mathcal{I}}^*_{1},\hdots,\bar{\mathcal{I}}^*_{2^{n-1}-1}$ form an IGC code 
set $\mathcal{I}(2^{n+p},2^n,2^m,2^{m-p})$.
\begin{IEEEproof}
 Please see Appendix B.
\end{IEEEproof}
\end{theorem}
\begin{corollary}\label{igcremark1}
In \textit{Theorem} \ref{thm2}, $g(\textbf{x},\textbf{u})=f(\textbf{x},\textbf{u}_{W'})+h(\textbf{u})$, where $f(\textbf{x},\textbf{u}_{W'})$
is given in (\ref{expgbf1}) with $\mathcal{G}_{\textbf{x}_{J_1}\textbf{x}_\mathcal{Z}}=0$ and $h(\textbf{u})$ is given in (\ref{expgbf2}). Therefore,
$G\left(g(\textbf{c},\textbf{u})\right)$ is a path over the vertices $u_0,u_1,\hdots,u_{n-1}$ for all $\textbf{c}\in \{0,1\}^m$. Hence,
by following the proof of \textit{Corollary} \ref{re2}, the maximum column sequence PMEPR of the IGC code set $\mathcal{I}(2^{n+p},2^n,2^m,2^{m-p})$
is $2$.
\end{corollary}
We have illustrated \textit{Theorem} \ref{thm2} in the below given example.
\begin{example}
 We consider $m=5,n=2, k=1,p=1, W=\{w_0\}=\{0\},\{w_1\}=\{1\}$. Therefore, $W'$ is a null set. We also consider 
Let $f:\{0,1\}^5\rightarrow \mathbb{Z}_4$ be a GBF given by
\begin{equation}
\begin{split}
f=2((1-x_0)(x_3x_2+x_2x_1)+x_0(x_1x_3+x_3x_2)\\
+x_0+2x_1+3x_3+x_4+2,
\end{split}
\end{equation}
and the function $h(u_0,u_1)$ is given by
\begin{equation}
 h(u_0,u_1)=2u_0u_1.
\end{equation}
From $f$, it is clear that 
$J_1=\{0\},J_1'=\{1,2,3\},$,  $\mathcal{Z}=\{4\}$, $\textbf{x}_{J_1}=(x_0)$, $\textbf{x}_\mathcal{Z}=(x_4)$ and $\mathcal{G}_{\textbf{x}_{J_1}\textbf{x}_\mathcal{Z}}=0$. 
From \textit{Theorem} \ref{thm2}, \textit{Corollary} \ref{igcremark1}, (\ref{igc1}), and (\ref{igc2}), we obtain $\mathcal{I}(8,4,32,16)$. 
The code groups $\mathcal{I}_0,\mathcal{I}_1,\bar{\mathcal{I}}^*_{0},\bar{\mathcal{I}}^*_{1}$ are given in TABLE II.
\begin{table*}[!htbp]
\begin{center}
\begin{tabular}{|l|l|l|l|}
\hline
\multicolumn{4}{|l|}{$\mathcal{I}_0$}                                                                                                                                                                                                                                                                                                                                                                                                                                                                                                                                                                                                                                                                                                                                                                                                                                                                                                                                                                                                                                                                                                                                                                                                                                                                                                              \\ \hline
$\psi(S_0)$       & \begin{tabular}[c]{@{}l@{}}$2   3   0   1    2   3    2   1    1    2   3  2   3   0   3    0   3   0    1   2   3   0  3   2    2   3   0    3   0    1   0   1$\\     $2   1   0   3    2   1    2   3    1    0   3  0   3   2   3    2   3   2    1   0   3   2  3   0    2   1   0    1   0    3   0   3$\\     $2   3   2   3    2   3    0   3    1    2   1  0   3   0   1    2   3   0    3   0   3   0  1   0    2   3   2    1   0    1   2   3$\\     $0   3   0   3    0   3    2   3    3    2   3  0   1   0   3    2   1   0    1   0   1   0  3   0    0   3   0    1   2    1   0   3$\end{tabular}                                 & $\psi(S_2)$       & \textbf{\begin{tabular}[c]{@{}l@{}}$2   3   0   1    2   3    2   1    1    2   3  2   3   0   3    0    1    2   3    0   1  2  1    0   0    1    2   1    2   3    2    3$\\    $ 2   1   0   3    2   1    2   3    1    0   3  0   3   2   3    2    1    0   3    2   1  0  1    2   0    3    2   3    2   1    2    1$\\    $ 2   3   2   3    2   3    0   3    1    2   1  0   3   0   1    2    1    2   1    2   1  2  3    2   0    1    0   3    2   3    0    1$\\    $ 0   3   0   3    0   3    2   3    3    2   3  0   1   0   3    2    3    2   3    2   3  2  1    2   2    1    2   3    0   3    2    1$\end{tabular}}     \\ \hline
\multicolumn{4}{|l|}{$\mathcal{I}_1$}                                                                                                                                                                                                                                                                                                                                                                                                                                                                                                                                                                                                                                                                                                                                                                                                                                                                                                                                                                                                                                                                                                                                                                                                                                                                                                              \\ \hline
$\psi(S_1)$       & \begin{tabular}[c]{@{}l@{}}$2    1   0    3    2    1    2    3    1   0   3  0   3    2   3   2   3    2    1    0   3    2  3    0    2    1   0   1   0   3   0    3$\\    $ 2    3   0    1    2    3    2    1    1   2   3  2   3    0   3   0   3    0    1    2   3    0  3    2    2    3   0   3   0   1   0    1$\\    $ 2    1   2    1    2    1    0    1    1   0   1  2   3    2   1   0   3    2    3    2   3    2  1    2    2    1   2   3   0   3   2    1 $\\    $ 0    1   0    1    0    1    2    1    3   0   3  2   1    2   3   0   1    2    1    2   1    2  3    2    0    1   0   3   2   3   0   1 $\end{tabular}    & $\psi(S_3)$       & \begin{tabular}[c]{@{}l@{}}$2    1   0    3    2    1    2    3    1   0   3  0   3    2   3   2    1   0   3   2    1   0  1   2   0   3    2    3    2   1    2   1$\\     $2    3   0    1    2    3    2    1    1   2   3  2   3    0   3   0    1   2   3   0    1   2  1   0   0   1    2    1    2   3    2   3$\\     $2    1   2    1    2    1    0    1    1   0   1  2   3    2   1   0    1   0   1   0    1   0  3   0   0   3    0    1    2   1    0   3$  \\     $0    1   0    1    0    1    2    1    3   0   3  2   1    2   3   0    3   0   3   0    3   0  1   0   2   3    2    1    0   1    2   3$\end{tabular}        \\ \hline
\multicolumn{4}{|l|}{$\bar{\mathcal{I}}^*_{0}$}                                                                                                                                                                                                                                                                                                                                                                                                                                                                                                                                                                                                                                                                                                                                                                                                                                                                                                                                                                                                                                                                                                                                                                                                                                                                                                    \\ \hline
$\psi(\bar(S)_0)$ & \begin{tabular}[c]{@{}l@{}}$3    0   1   2    1    0    3    0   2    1    2  3   2   3    2   3    0    1    2   3   2    1  0    1   3   2    3    0   3    0    3    $0\\   $ 1    0   3   2    3    0    1    0   0    1    0  3   0   3    0   3    2    1    0   3   0    1  2    1   1   2    1    0   1    0    1    0$\\  $  3    0   3   0    1    0    1    2   2    1    0  1   2   3    0   1    0    1    0   1   2    1  2    3   3   2    1    2   3    0    1   2$ \\   $ 3    2   3   2    1    2    1    0   2    3    0  3   2   1    0   3    0    3    0   3   2    3  2    1   3   0    1    0   3    2    1   0$\end{tabular} & $\psi(\bar(S)_2)$ & \begin{tabular}[c]{@{}l@{}}$1   2    3    0   3   2   1   2    0   3   0  1    0    1   0    1    0    1    2   3   2    1  0    1   3   2    3    0   3    0    3    0$\\    $ 3   2    1    0   1   2   3   2    2   3   2  1    2    1   2    1    2    1    0   3   0    1  2    1   1   2    1    0   1    0    1    0$\\     $1   2    1    2   3   2   3   0    0   3   2  3    0    1   2    3    0    1    0   1   2    1  2    3   3   2    1    2   3    0    1    2$ \\     $1   0    1    0   3   0   3   2    0   1   2  1    0    3   2    1    0    3    0   3   2    3  2    1   3   0    1    0   3    2    1    0$\end{tabular} \\ \hline
\multicolumn{4}{|l|}{$\bar{\mathcal{I}}^*_{1}$}                                                                                                                                                                                                                                                                                                                                                                                                                                                                                                                                                                                                                                                                                                                                                                                                                                                                                                                                                                                                                                                                                                                                                                                                                                                                                                    \\ \hline
$\psi(\bar(S)_1)$ & \begin{tabular}[c]{@{}l@{}}$1    0    3   2   3    0   1    0    0    1   0 3    0   3   0   3   2    1   0   3    0    1 2    1    1   2   1    0    1    0   1    0$\\     $3    0    1   2   1    0   3    0    2    1   2 3    2   3   2   3   0    1   2   3    2    1 0    1    3   2   3    0    3    0   3    0 $ \\     $1    0    1   0   3    0   3    2    0    1   2 1    0   3   2   1   2    1   2   1    0    1 0    3    1   2   3    2    1    0   3    2 $\\     $1    2    1   2   3    2   3    0    0    3   2 3    0   1   2   3   2    3   2   3    0    3 0    1    1   0   3    0    1    2   3    0 $\end{tabular}         & $\psi(S_3)$       & \begin{tabular}[c]{@{}l@{}}$3   2   1    0    1   2    3   2   2   3    2 1   2    1    2    1   2    1   0   3    0    1 2    1    1   2   1    0    1    0   1    0$\\    $1   2   3    0    3   2    1   2   0   3    0 1   0    1    0    1   0    1   2   3    2    1 0    1    3   2   3    0    3    0   3    0$ \\    $3   2   3    2    1   2    1   0   2   3    0 3   2    1    0    3   2    1   2   1    0    1 0    3    1   2   3    2    1    0   3    2$\\    $3   0   3    0    1   0    1   2   2   1    0 1   2    3    0    1   2    3   2   3    0    3 0    1    1   0   3    0    1    2   3    0$\end{tabular}            \\ \hline
\end{tabular}
\end{center}
\caption{$\mathcal{I}(8,4,32,16)$ over the alphabet $\mathbb{Z}_4$ with maximum column sequence PMEPR $2$.}
\end{table*}
\begin{figure}[!h]\label{gig2}
\centering
\includegraphics[height=6cm,width=8cm]{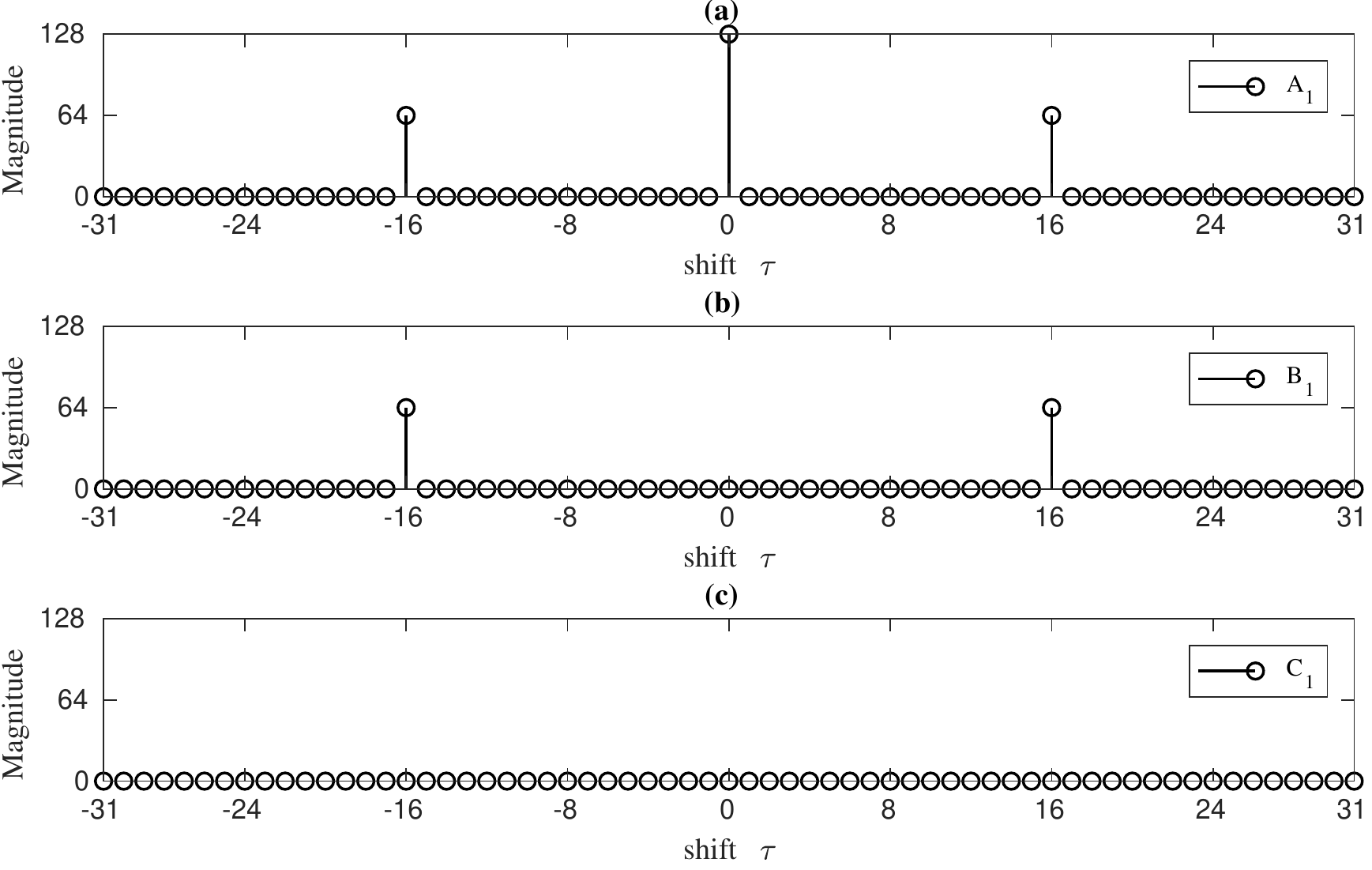}
\caption{Correlation plots of $\mathcal{I}(8,4,32,16)$ in TABLE II.}
%\caption{The graph of the quadratic form $x_2x_3+x_3x_1+x_0x_2+x_0x_3+x_0x_1+x_0x_4$.}
\end{figure}
Fig. 2-(a) represents AACF of any code given in TABLE II. Fig. 2-(b)  represents AACF of between two distinct
codes from same code group and Fig. 2-(c)  represents AACF of between two 
codes from different code groups in TABLE II.
\end{example}
\section{Comparison of the Proposed Construction With Existing ZCCS and IGC code set constructions}
In this section, we compare our proposed ZCCS construction with the construction given in \cite{p_tcom,chen_zccs,li_ZCCS} and the proposed IGC 
code set construction with \cite{sarkar_igc,jli_igc_2008}.

The constructions of ZCCS given in \cite{p_tcom} and \cite{chen_zccs} both are based on second-order GBFs but the maximum 
column sequence PMEPR depends on the number of subcarriers or the number of sequences in the code. In order to increase number of 
users in a ZCCS based MC-CDMA system with large ZCZ, we need to increase number of subcarriers which increase column sequence
PMEPR. For this scenario all GBFs based ZCCS degrades the performance of a ZCCS based MC-CDMA system.
%Therefore, the large number of sub-carrier in a MC-CDMA
%system degrades the performance of a existing ZCCS based
%MC-CDMA system. 
Our proposed constrcution of both ZCCS and IGC code set are based on higher-order ($\geq 2$) GBFs with maximum column sequence PMEPR $2$.
Additionaly, we have linked our proposed ZCCS with higher order RM code unlike the order of GBFs used in \cite{p_tcom,chen_zccs}.
Although, the constrcution given in \cite{li_ZCCS} can generate ZCCS with maximum column sequence PMEPR $2$ but the 
constrcution is based on Golay sequences with
large zero autocorrelation zone and orthogonal matrix. Therefore, the constrcution given in \cite{li_ZCCS} may not be suitable for fast 
hardware generation specially for long ZCCS \cite{uda2014}.

The constrcution of IGC code set given in \cite{sarkar_igc} is based on second-order Boolean function but maximum column sequence PMEPR
depends on the number of constituent sequences in the code. Therefore, in a large subcarrier MC-CDMA system, the IGC code set obtained from
\cite{sarkar_igc} cannot provide a tight PMEPR upper bound unlike the proposed IGC code set. The IGC code set construction given in \cite{jli_igc_2008} is 
based on CCCs and orthogonal matrix as well as the maximum column sequence PMEPR depends on the number of constituent sequences in the code. 
Therefore, the IGC code set obtained from \cite{jli_igc_2008} may not be suitable for fast hardware generation as well as for a large 
subcarrier MC-CDMA system where the high PMEPR value may not be acceptable. Hence, our proposed construction is more suitable than the above 
mentioned constructions. We also have provided a comparison table for column sequence PMEPR for ZCCS and IGC code set in TABLE III. 
\begin{table}[!t]
\small
\resizebox{\textwidth}{!}{
\begin{tabular}{|l|l|l|l|l|}
\hline
         Code Set     & Method                                                                                                                         & PMEPR& Remark \\ \hline
         ZCCS \cite{p_tcom}         & \begin{tabular}[c]{@{}l@{}}Based on \\second-order\\ GBFs\end{tabular}                                                        & $> 2$& Direct     \\ \hline
      ZCCS \cite{chen_zccs}         & \begin{tabular}[c]{@{}l@{}}Based on \\ second-order\\ GBFs\end{tabular}                                                        & $>2$&Direct     \\ \hline
       ZCCS \cite{li_ZCCS}          & \begin{tabular}[c]{@{}l@{}}Golay sequences \\ with large zero \\ autocorrelation \\ zone and \\orthogonal \\ matrix\end{tabular} & $2$&Indirect     \\ \hline
  IGC \cite{jli_igc_2008} & \begin{tabular}[c]{@{}l@{}}Based on\\ CCCs and\\ orthogonal \\ matrix\end{tabular}                                        & $>2 $&Indirect     \\ \hline
     IGC \cite{sarkar_igc} & \begin{tabular}[c]{@{}l@{}}Based on \\ second-order\\ GBFs\end{tabular}                                                        & $>2 $&Direct     \\ \hline
Proposed                                ZCCS         & \begin{tabular}[c]{@{}l@{}}Based on \\ higher-order ($\geq 2$)\\ GBFs\end{tabular}                                & $2$&Direct     \\ \hline
Proposed                                IGC & \begin{tabular}[c]{@{}l@{}}Based on \\ higher-order ($\geq 2$)\\ GBFs\end{tabular}                                & $2$&Direct     \\ \hline
\end{tabular}}
\caption{Comparison of proposed construction with \cite{p_tcom,chen_zccs,li_ZCCS,jli_igc_2008,sarkar_igc} }
\end{table}
\section{CONCLUSION}
\label{sec:con}
This paper focuses on 
a direct construction of ZCCS and IGC code set with maximum column sequence PMEPR $2$. Both the constructions are based on GBFs of order no less than 
$2$. We also have linked our proposed construction with graph. The construction of ZCCS achieves the theoritical upper bound. The
maximum column sequence PMEPR of 
existed ZCCS, based on GBFs depend on number of constituent sequence in a code. Also, the maximum column sequence PMEPR of all existed IGC code
sets depend on the number of constituent sequences in a code.
%Our proposed construction is direct and also we have optimized the maximum column 
%sequence PMEPR to $2$ for both ZCCS and IGC code set.
%using graphical representation of second-order RM codes. The proposed
%construction valids for any number of isolated vertices present in the graph, is capable of generating
%optimal ZCCS with respect to the set size upper bound in \textit{Lemma} \ref{lemma5}. It is noted that the construction of CCCs in \cite{arthina} is
%a special case of our work work when the number of isolated vertices is set to zero. Flexible ZCZ
%width and set size can be obtained by varying the number of isolated vertices.
\begin{appendices}
%\section{Proof of \textnormal{\textit{Lemma} \ref{lemma6}}}
\section{Proof of \textnormal{\textit{Theorem} \ref{thm1}}}
To prove \textit{Theorem} \ref{thm1}, it is enough to show that the AACF of any code from the set given in 
(\ref{mainset}) is zero for all nonzero time shifts inside the ZCZ, $2^{m-p}$ and the ACCF of any two codes is zero for all time shifts inside the ZCZ, $2^{m-p}$. We define the following binary vectors:
$\textbf{t}_1=(t_0,t_1,\hdots,t_{k-1}), \textbf{t}_2=(t_k,t_{k+1},\hdots,t_{n-2})$, and $\textbf{t}_3=(t_{n-1},t_n,\hdots,t_{n+p-2})$, where $(t_0,t_1,\hdots,t_{n+p-2})$ is the binary representation of the integer $t$ which already define in \textit{Theorem} \ref{thm1}. 
Also, we define $\textbf{t}'_1=(t'_0,t'_1,\hdots,t'_{k-1}), \textbf{t}'_2=(t'_k,t'_{k+1},\hdots,t'_{n-2})$, and $\textbf{t}'_3=(t'_{n-1},t'_n,\hdots,t'_{n+p-2})$, where $(t'_0,t'_1,\hdots,t'_{n+p-2})$ is the binary representation of the integer $t'$. In the expression of $S_t$ and $\bar{S}_t$, we assume 
\begin{equation}\label{peq1}
\begin{split}
\eta_1&=\frac{q}{2}\left(\displaystyle\sum_{\alpha=0}^{k-1}u_{w_\alpha}x_{j_\alpha}+u_{w_k}x_\gamma\right)\\
      &=\frac{q}{2}\left(\textbf{u}_{W}\cdot \textbf{x}_{J_1}+u_{w_k}x_\gamma\right),\\
\eta_2&=\frac{q}{2}\left(\displaystyle\sum_{\alpha=0}^{k-1}u_{w_\alpha}\bar{x}_{j_\alpha}+\bar{u}_{w_k}x_\gamma\right)\\
      &=\frac{q}{2}\left(\textbf{u}_{W}\cdot\bar{\textbf{x}}_{J_1}+\bar{u}_{w_k}x_\gamma\right),
\end{split}
\end{equation},
where $\bar{\textbf{x}}_{J_1}=\textbf{1}_k-\textbf{x}_{J_1}$ and $\bar{u}_{w_k}=1-u_{w_k}$.
Let us start with $\mathscr{C}(\psi(S_t),\psi(S_{t'}))(\tau)$, where the expression denotes the ACCF of the codes $\psi(S_t)$ and $\psi(S_{t'})$ at the time shift $\tau$ when $t\neq t'$, and AACF of $\psi(S_t)$ (or, $\psi(S_{t'})$ ) at the time shift $\tau$ when $t=t'$. The expression $\mathscr{C}(\psi(S_t),\psi(S_{t'}))(\tau)$ can be written as  
\begin{equation}\label{eq1}
\begin{split}
&\mathscr{C}(\psi(S_t),\psi(S_{t'}))(\tau)\\
&=\displaystyle\sum_{\textbf{u}\in\{0,1\}^n}\mathscr{C}\left(g+\eta_1+\frac{q}{2}\left(\textbf{t}_1\cdot\textbf{x}_{J_1}+\textbf{t}_2\cdot \textbf{u}_{W'}+\textbf{t}_3\cdot\textbf{x}_\mathcal{Z}\right),\right.\\ 
&~~~~~~~~~~~~~\left.g+\eta_1+\frac{q}{2}\left(\textbf{t}'_1\cdot\textbf{x}_{J_1}+\textbf{t}'_2\cdot \textbf{u}_{W'}+\textbf{t}'_3\cdot\textbf{x}_\mathcal{Z}\right)\right)(\tau)\\
&=\sum_{\textbf{u}_{W'}=\textbf{e}}(-1)^{(\textbf{t}_2-\textbf{t}'_2)\cdot \textbf{e}}\\ &~~~~~\times\left( \sum_{\textbf{u}_W,u_{w_k}}\mathscr{C}\left(f\arrowvert_{\textbf{u}_{W'}=\textbf{e}}+\eta_1+\frac{q}{2}\left(\textbf{t}_1\cdot\textbf{x}_{J_1}+\textbf{t}_3\cdot\textbf{x}_\mathcal{Z}\right),\right.\right.\\ 
&~~~~~~~~~~~~~\left.\left.f\arrowvert_{\textbf{u}_{W'}=\textbf{e}}+\eta_1+\frac{q}{2}\left(\textbf{t}'_1\cdot\textbf{x}_{J_1}+\textbf{t}'_3\cdot\textbf{x}_\mathcal{Z}\right)\right)(\tau)\right)\\
&=\sum_{\textbf{u}_{W'}=\textbf{e}}(-1)^{(\textbf{t}_2-\textbf{t}'_2)\cdot \textbf{e}} \mathcal{A},
\end{split}
\end{equation}
where
\begin{equation}\label{eq2}
\begin{split}
\mathcal{A}&=\left( \sum_{\textbf{u}_W,u_{w_k}}\mathscr{C}\left(f\arrowvert_{\textbf{u}_{W'}=\textbf{e}}+\eta_1+\frac{q}{2}\left(\textbf{t}_1\cdot\textbf{x}_{J_1}+\textbf{t}_3\cdot\textbf{x}_\mathcal{Z}\right),\right.\right.\\ 
&~~~~~~\left.\left.f\arrowvert_{\textbf{u}_{W'}=\textbf{e}}+\eta_1+\frac{q}{2}\left(\textbf{t}'_1\cdot\textbf{x}_{J_1}+\textbf{t}'_3\cdot\textbf{x}_\mathcal{Z}\right)\right)(\tau)\right).
\end{split}
\end{equation}
Now, $\mathcal{A}$ given in (\ref{eq2}), can be expressed as
\begin{equation}\label{eqqq2}
\begin{split}
\mathcal{A}=\mathcal{A}_1+\mathcal{A}_2,
\end{split}
\end{equation}
where 
\begin{equation}\label{eq3}
\begin{split}
\mathcal{A}_1&=\left( \sum_{\textbf{u}_W,u_{w_k}}\sum_{\textbf{c}_1\neq\textbf{c}_2}\mathscr{C}\left(f\arrowvert_{\textbf{u}_{W'}=\textbf{e},\textbf{x}_{J_1}=\textbf{c}_1}+\eta_1\arrowvert_{\textbf{x}_{J_1}=\textbf{c}_1}\right.\right.\\&~~~~~~\left.\left.+\frac{q}{2}\left(\textbf{t}_1\cdot\textbf{c}_1+\textbf{t}_3\cdot\textbf{x}_\mathcal{Z}\right),
f\arrowvert_{\textbf{u}_{W'}=\textbf{e},\textbf{x}_{J_1}=\textbf{c}_2}+\eta_1\arrowvert_{\textbf{x}_{J_1}=\textbf{c}_2}\right.\right. \\ &~~~~~~\left.\left.+\frac{q}{2}\left(\textbf{t}'_1\cdot\textbf{c}_2+\textbf{t}'_3\cdot\textbf{x}_\mathcal{Z}\right)\right)(\tau)\right),
\end{split}
\end{equation}
\begin{equation}\label{eq4}
\begin{split}
\mathcal{A}_2&=\left( \sum_{\textbf{u}_W,u_{w_k}}\sum_{\textbf{c}\in\{0,1\}^k}\mathscr{C}\left(f\arrowvert_{\textbf{u}_{W'}=\textbf{e},\textbf{x}_{J_1}=\textbf{c}}+\eta_1\arrowvert_{\textbf{x}_{J_1}=\textbf{c}}\right.\right.\\&~~~~~\left.\left.+\frac{q}{2}\left(\textbf{t}_1\cdot\textbf{c}+\textbf{t}_3\cdot\textbf{x}_\mathcal{Z}\right),
f\arrowvert_{\textbf{u}_{W'}=\textbf{e},\textbf{x}_{J_1}=\textbf{c}}+\eta_1\arrowvert_{\textbf{x}_{J_1}=\textbf{c}}\right.\right. \\ &~~~~~~\left.\left.+\frac{q}{2}\left(\textbf{t}'_1\cdot\textbf{c}+\textbf{t}'_3\cdot\textbf{x}_\mathcal{Z}\right)\right)(\tau)\right).
\end{split}
\end{equation}
From (\ref{peq1}) and (\ref{eq3}), we have
\begin{equation}\label{eq5}
\begin{split}
\mathcal{A}_1&=\left( \sum_{u_{w_k}}\sum_{\textbf{c}_1\neq\textbf{c}_2}\mathscr{C}\left(f\arrowvert_{\textbf{u}_{W'}=\textbf{e},\textbf{x}_{J_1}=\textbf{c}_1}+\frac{q}{2}u_{w_k}\right.\right.\\&~~~~~\left.\left.+\frac{q}{2}\left(\textbf{t}_1\cdot\textbf{c}_1+\textbf{t}_3\cdot\textbf{x}_\mathcal{Z}\right), 
f\arrowvert_{\textbf{u}_{W'}=\textbf{e},\textbf{x}_{J_1}=\textbf{c}_2}+\frac{q}{2}u_{w_k}\right.\right. \\ &~~~~~\left.\left.+\frac{q}{2}\left(\textbf{t}'_1\cdot\textbf{c}_2+\textbf{t}'_3\cdot\textbf{x}_\mathcal{Z}\right)\right)(\tau)\sum_{\textbf{u}_W}(-1)^{\textbf{u}_W\cdot(\textbf{c}_1-\textbf{c}_2)}\right).
\end{split}
\end{equation}
Since, $\textbf{c}_1\neq \textbf{c}_2$, $\sum_{\textbf{u}_W}(-1)^{\textbf{u}_W\cdot(\textbf{c}_1-\textbf{c}_2)}=0$ and therefore,
from (\ref{eq3})
\begin{equation}\label{eq6}
\mathcal{A}_1=0~\forall ~\tau.
\end{equation}
From (\ref{peq1}) and (\ref{eq4}), we have 
\begin{equation}\label{neq1}
\begin{split}
\mathcal{A}_2&=\left( \sum_{u_{w_k}}\sum_{\textbf{c}\in\{0,1\}^k}2^k(-1)^{(\textbf{t}_1-\textbf{t}'_1)\cdot\textbf{c}}\mathscr{C}\left(f\arrowvert_{\textbf{u}_{W'}=\textbf{e},\textbf{x}_{J_1}=\textbf{c}}+\right.\right.\\&~~~~~~\left.\left.\frac{q}{2}u_{w_k}x_{\gamma_\textbf{c}}+\frac{q}{2}\left(\textbf{t}_3\cdot\textbf{x}_\mathcal{Z}\right), 
f\arrowvert_{\textbf{u}_{W'}=\textbf{e},\textbf{x}_{J_1}=\textbf{c}}+\frac{q}{2}u_{w_k}x_{\gamma_\textbf{c}}\right.\right. \\ &~~~~~~~~~~~~~~~~~~~~~~~~~~~~~~~~\left.\left.+\frac{q}{2}\left(\textbf{t}'_3\cdot\textbf{x}_\mathcal{Z}\right)\right)(\tau)\right).
\end{split}
\end{equation}
Assume, $\mathcal{F}\arrowvert_{\textbf{x}_{J_1}=\textbf{c}}=f\arrowvert_{\textbf{u}_{W'}=\textbf{e},\textbf{x}_{J_1}=\textbf{c}}$. 
The following expression present in (\ref{eq6}), can be expressed as follows:
\begin{equation}\label{teq}
 \begin{split}
 \mathscr{C}&\left(f\arrowvert_{\textbf{u}_{W'}=\textbf{e},\textbf{x}_{J_1}=\textbf{c}}+\frac{q}{2}u_{w_k}x_{\gamma_\textbf{c}}+\frac{q}{2}\left(\textbf{t}_3\cdot\textbf{x}_\mathcal{Z}\right),\right.\\ 
&~~~~~~~~~~~~\left.\left.f\arrowvert_{\textbf{u}_{W'}=\textbf{e},\textbf{x}_{J_1}=\textbf{c}}+\frac{q}{2}u_{w_k}x_{\gamma_\textbf{c}}+\frac{q}{2}\left(\textbf{t}'_3\cdot\textbf{x}_\mathcal{Z}\right)\right)(\tau)\right) \\
&=\displaystyle\sum_{\textbf{d}'\neq \textbf{d}''}(-1)^{(\textbf{t}_3\cdot\textbf{d}'-\textbf{t}'_3\cdot\textbf{d}'')}\mathscr{C}\left(\mathcal{F}\arrowvert_{\textbf{x}_{J_1}\textbf{x}_\mathcal{Z}=\textbf{c}\textbf{d}'}+\frac{q}{2}u_{w_k}x_{\gamma_\textbf{c}},\right.\\ 
&~~~~~~~~~~~~~~\left.\left.\mathcal{F}\arrowvert_{\textbf{x}_{J_1}\textbf{x}_\mathcal{Z}=\textbf{c}\textbf{d}''}+\frac{q}{2}u_{w_k}x_{\gamma_\textbf{c}}\right)(\tau)\right)\\
&~~+\sum_{\textbf{d}'=\textbf{d}''}(-1)^{(\textbf{t}_3\cdot\textbf{d}'-\textbf{t}'_3\cdot\textbf{d}'')}\mathscr{C}\left(\mathcal{F}\arrowvert_{\textbf{x}_{J_1}\textbf{x}_\mathcal{Z}=\textbf{c}\textbf{d}'}+\frac{q}{2}u_{w_k}x_{\gamma_\textbf{c}},\right.\\ 
&~~~~~~~~~~~~~~\left.\left.\mathcal{F}\arrowvert_{\textbf{x}_{J_1}\textbf{x}_\mathcal{Z}=\textbf{c}\textbf{d}''}+\frac{q}{2}u_{w_k}x_{\gamma_\textbf{c}}\right)(\tau)\right)\\
&=\displaystyle\sum_{\textbf{d}'\neq \textbf{d}''}(-1)^{(\textbf{t}_3\cdot\textbf{d}'-\textbf{t}'_3\cdot\textbf{d}'')}\mathscr{C}\left(\mathcal{F}\arrowvert_{\textbf{x}_{J_1}\textbf{x}_\mathcal{Z}=\textbf{c}\textbf{d}'}+\frac{q}{2}u_{w_k}x_{\gamma_\textbf{c}},\right.\\ 
&~~~~~~~~~~~~~~\left.\left.\mathcal{F}\arrowvert_{\textbf{x}_{J_1}\textbf{x}_\mathcal{Z}=\textbf{c}\textbf{d}''}+\frac{q}{2}u_{w_k}x_{\gamma_\textbf{c}}\right)(\tau)\right)\\
&~~+\sum_{\textbf{d}'}(-1)^{(\textbf{t}_3-\textbf{t}'_3)\cdot\textbf{d}'}\mathscr{A}\left(\mathcal{F}\arrowvert_{\textbf{x}_{J_1}\textbf{x}_\mathcal{Z}=\textbf{c}\textbf{d}'}+\frac{q}{2}u_{w_k}x_{\gamma_\textbf{c}}\right)(\tau).
 \end{split}
\end{equation}
The only term associated with restricted vertices $x_{j_0},x_{j_1},\hdots,x_{j_{k-1}}$ and isolated vertices $x_{m-p},x_{m-p+1},\hdots,x_{m-1}$
can be expressed as follows:
\begin{equation}\nonumber
 \begin{split}
  \mathcal{G}_{\textbf{x}_{J_1}\textbf{x}_\mathcal{Z}}\!\!\!=\!\!\!\sum_{r=1}^k\sum_{\alpha=0}^{p-1}\sum_{0\leq \alpha_0<\alpha_1<\hdots<\alpha_{r}< k}
  \!\!\!\!\!\!\!\!\!\!\!\!\!\!\!\!\!\!
\varrho^{m-p+\alpha}_{j_0,j_1,\hdots,j_r}x_{j_{\alpha_0}}x_{j_{\alpha_1}}\!\!\!\!\!\hdots x_{j_{\alpha_r}}x_{m-p+\alpha},
 \end{split}
\end{equation}
where, we have introduced the term $\mathcal{G}_{\textbf{x}_{J_1}\textbf{x}_\mathcal{Z}}$ in (\ref{teq1}). 
By taking sum over $u_{w_k}$ and then using \textit{Lemma} \ref{lemmaz}, the following expression of (\ref{teq}) can be expressed as follows:
\begin{equation}\label{teq2}
 \begin{split}
  &\sum_{u_{w_k}}\mathscr{C}\left(\mathcal{F}\arrowvert_{\textbf{x}_{J_1}\textbf{x}_\mathcal{Z}=\textbf{c}\textbf{d}'}+\frac{q}{2}u_{w_k}x_{\gamma_\textbf{c}},\right.\\ 
&~~~~~~~~~~~~~~\left.\left.\mathcal{F}\arrowvert_{\textbf{x}_{J_1}\textbf{x}_\mathcal{Z}=\textbf{c}\textbf{d}''}+\frac{q}{2}u_{w_k}x_{\gamma_\textbf{c}}\right)(\tau)\right)\\
&=\begin{cases}
\omega^{\mathcal{G}_{\textbf{c}\textbf{d}'-{\textbf{c}\textbf{d}''}}}\omega^{(d_1'-d_1'')g_{m-p}+\hdots+(d_p'-d_p'')g_{m-1}}2^{m-(k+p)+1},  \\
~~~~~~~~~~\tau=(d_1'-d_1'')2^{m-p}+\hdots+(d_p'-d_p'')2^{m-1},\\
0, ~~~~~~~~~~~~\textnormal{otherwise},
\end{cases}\\
&=\begin{cases}
\omega^{\mathcal{G}_{\textbf{c}\textbf{d}'}-\mathcal{G}_{\textbf{c}\textbf{d}''}}\omega^{(\textbf{d}'-\textbf{d}'')\cdot (g_{m-p},\hdots,g_{m-1})}2^{m-(k+p)+1},  \\
~~~~~~~~~~\tau=(\textbf{d}'-\textbf{d}'')\cdot (2^{m-p},\hdots,2^{m-1}),\\
0, ~~~~~~~~~~~~\textnormal{otherwise}.
 \end{cases}
 \end{split}
\end{equation}
In the above expression $\textbf{d}'$ and $\textbf{d}''$ are two $p$-length binary vectors 
and $(\textbf{d}'-\textbf{d}'')\cdot (2^{m-p},\hdots,2^{m-1})$ takes the value $\tau$. It is possible to get another pair of vectors $\textbf{d}'$
and $\textbf{d}''$ such that $(\textbf{d}'-\textbf{d}'')\cdot (2^{m-p},\hdots,2^{m-1})$ also takes the value $\tau$. We assume, for all possible $\textbf{d}'$ and $\textbf{d}''$ in 
$\{0,1\}^p$, $(\textbf{d}'-\textbf{d}'')\cdot (2^{m-p},\hdots,2^{m-1})$ takes the integer values $\tau_1,\tau_2,\hdots,\tau_\varsigma$, where 
$0\leq \varsigma<3^p$.
Therefore,
we define $\mathcal{K}_i=\left\{(\textbf{d}',\textbf{d}''): \textbf{d}'\neq \textbf{d}'', (\textbf{d}'-\textbf{d}'')\cdot (2^{m-p},\hdots,2^{m-1})=\tau_i \right\}$, for 
$i=1,2,\hdots,\varsigma$. From (\ref{teq}) and (\ref{teq2}), we have
\begin{equation}\label{teq3}
 \begin{split}
&\sum_{u_{w_k}}\mathscr{C}\left(f\arrowvert_{\textbf{u}_{W'}=\textbf{e},\textbf{x}_{J_1}=\textbf{c}}+\frac{q}{2}u_{w_k}x_{\gamma_\textbf{c}}+\frac{q}{2}\left(\textbf{t}_3\cdot\textbf{x}_\mathcal{Z}\right),\right.\\ 
&~~~~~~~~~~~~\left.\left.f\arrowvert_{\textbf{u}_{W'}=\textbf{e},\textbf{x}_{J_1}=\textbf{c}}+\frac{q}{2}u_{w_k}x_{\gamma_\textbf{c}}+\frac{q}{2}\left(\textbf{t}'_3\cdot\textbf{x}_\mathcal{Z}\right)\right)(\tau)\right) \\
&=\displaystyle\sum_{\textbf{d}'\neq \textbf{d}''}(-1)^{(\textbf{t}_3\cdot\textbf{d}'-\textbf{t}'_3\cdot\textbf{d}'')}\mathscr{C}\left(\mathcal{F}\arrowvert_{\textbf{x}_{J_1}\textbf{x}_\mathcal{Z}=\textbf{c}\textbf{d}'}+\frac{q}{2}u_{w_k}x_{\gamma_\textbf{c}},\right.\\ 
&~~~~~~~~~~~~~~\left.\left.\mathcal{F}\arrowvert_{\textbf{x}_{J_1}\textbf{x}_\mathcal{Z}=\textbf{c}\textbf{d}''}+\frac{q}{2}u_{w_k}x_{\gamma_\textbf{c}}\right)(\tau)\right)\\
&=\displaystyle\sum_{i=1}^\varsigma\sum_{(\textbf{d}',\textbf{d}'')\in \mathcal{K}_i}(-1)^{(\textbf{t}_3\cdot\textbf{d}'-\textbf{t}'_3\cdot\textbf{d}'')}\mathscr{C}\left(\mathcal{F}\arrowvert_{\textbf{x}_{J_1}\textbf{x}_\mathcal{Z}=\textbf{c}\textbf{d}'}+\frac{q}{2}u_{w_k}x_{\gamma_\textbf{c}},\right.\\ 
&~~~~~~~~~~~~~~\left.\left.\mathcal{F}\arrowvert_{\textbf{x}_{J_1}\textbf{x}_\mathcal{Z}=\textbf{c}\textbf{d}''}+\frac{q}{2}u_{w_k}x_{\gamma_\textbf{c}}\right)(\tau)\right)\\
&=\begin{cases}
\displaystyle\sum_{i=1}^\varsigma\sum_{(\textbf{d}',\textbf{d}'')\in \mathcal{K}_i}(-1)^{(\textbf{t}_3\cdot\textbf{d}'-\textbf{t}'_3\cdot\textbf{d}'')}\omega^{\mathcal{G}_{\textbf{c}\textbf{d}'}-\mathcal{G}_{\textbf{c}\textbf{d}''}}\\\times\omega^{(\textbf{d}'-\textbf{d}'')\cdot (g_{m-p},\hdots,g_{m-1})}2^{m-(k+p)+1},  \\
~~~~~~~~~~\tau=(\textbf{d}'-\textbf{d}'')\cdot (2^{m-p},\hdots,2^{m-1}),\\
0, ~~~~~~~~~~~~\textnormal{otherwise}
\end{cases}\\
&=\begin{cases}
\displaystyle\sum_{(\textbf{d}',\textbf{d}'')\in \mathcal{K}_i}(-1)^{(\textbf{t}_3\cdot\textbf{d}'-\textbf{t}'_3\cdot\textbf{d}'')}\omega^{\mathcal{G}_{\textbf{c}\textbf{d}'}-\mathcal{G}_{\textbf{c}\textbf{d}''}}\\\times\omega^{(\textbf{d}'-\textbf{d}'')\cdot (g_{m-p},\hdots,g_{m-1})}2^{m-(k+p)+1},  \\
~~~~~~~~~~\tau=\tau_i, i=1,2,\hdots,\varsigma\\
0, ~~~~~~~~~~~~\textnormal{otherwise}.
 \end{cases}
 \end{split}
\end{equation}
For each of $\textbf{c}\in\{0,1\}^k$,  $G\left(\mathcal{F}\arrowvert_{\textbf{x}_{J_1}\textbf{x}_\mathcal{Z}=\textbf{c}\textbf{d}''}\right)$ is a path
over the vertices specified in $J_1'$. Therefore,
\begin{equation}\label{teq4}
 \begin{split}
  \sum_{u_{w_k}}\mathscr{A}\left(\mathcal{F}\arrowvert_{\textbf{x}_{J_1}\textbf{x}_\mathcal{Z}=\textbf{c}\textbf{d}'}+\frac{q}{2}u_{w_k}x_{\gamma_\textbf{c}}\right)(\tau)
  =\begin{cases}
    2^{m-k-p+1},~\tau=0,\\
    0, ~~~~~~~~~\textnormal{othewise}.
   \end{cases}
 \end{split}
\end{equation}
Therefore, from (\ref{neq1}), (\ref{teq}), (\ref{teq3}), and (\ref{teq4}), we have 
\begin{equation}\label{teq5}
\begin{split}
\mathcal{A}_2&=\left( \sum_{u_{w_k}}\sum_{\textbf{c}\in\{0,1\}^k}2^k(-1)^{(\textbf{t}_1-\textbf{t}'_1)\cdot\textbf{c}}\mathscr{C}\left(f\arrowvert_{\textbf{u}_{W'}=\textbf{e},\textbf{x}_{J_1}=\textbf{c}}+\right.\right.\\&~~~~~~~~~~~~~~~~~~~~~~~~~~~~~~~~\left.\frac{q}{2}u_{w_k}x_{\gamma_\textbf{c}}+\frac{q}{2}\left(\textbf{t}_3\cdot\textbf{x}_\mathcal{Z}\right),\right.\\ 
&~~~~~~~~~~~~~~~~~~~~~~~~~~\left.f\arrowvert_{\textbf{u}_{W'}=\textbf{e},\textbf{x}_{J_1}=\textbf{c}}+\frac{q}{2}u_{w_k}x_{\gamma_\textbf{c}}\right. \\ &~~~~~~~~~~~~~~~~~~~~~~~~~~~~~~~~\left.\left.+\frac{q}{2}\left(\textbf{t}'_3\cdot\textbf{x}_\mathcal{Z}\right)\right)(\tau)\right)\\
&=\begin{cases}
  2^{m-p+1}\displaystyle\sum_{\textbf{c}\in\{0,1\}^k}(-1)^{(\textbf{t}_1-\textbf{t}'_1)\cdot\textbf{c}}\\
  ~~~~~\times\displaystyle\sum_{\textbf{d}'\in\{0,1\}^p}
  (-1)^{(\textbf{t}_3-\textbf{t}_3')\cdot \textbf{d}'},~\tau=0\\
  2^{m-p+1}\displaystyle\sum_{(\textbf{d}',\textbf{d}'')\in \mathcal{K}_i}(-1)^{(\textbf{t}_3\cdot\textbf{d}'-\textbf{t}'_3\cdot\textbf{d}'')}\omega^{(\textbf{d}'-\textbf{d}'')\cdot (g_{m-p},\hdots,g_{m-1})}
  \\
  ~~~~~~\times\displaystyle\sum_{\textbf{c}\in\{0,1\}^k}(-1)^{(\textbf{t}_1-\textbf{t}'_1)\cdot\textbf{c}}\omega^{\mathcal{G}_{\textbf{c}\textbf{d}'}-\mathcal{G}_{\textbf{c}\textbf{d}''}},  \\
~~~~~~~~~~~~~~~~~~~~~~~~~~~~~~~~~~~~~\tau=\tau_i, i=1,2,\hdots,\varsigma\\
0, ~~~~~~~~~~~~~~~~~~~~~~~~~~~~~~~~~~~\textnormal{otherwise.}
 \end{cases}
\end{split}
\end{equation}
From (\ref{eq1}), (\ref{eqqq2}), and (\ref{eq3}), we have
%
%For all $\textbf{c}\in\{0,1\}^k$ and $\textbf{e}\in\{0,1\}^{n-k-1}$, $G(f\arrowvert_{\textbf{u}_{W'}=\textbf{e},\textbf{x}_{J_1}=\textbf{c}})$ contains a path over $m-k-p$ vertices specified in $J'_1$ and $p$ isolated vertices specified in $Z$. Therefore,
%by using \textit{Lemma} \ref{lm1} in (\ref{eq4}), we have
%\begin{equation}\label{feq1}
%\begin{split}
%\mathcal{A}_2=\begin{cases}
%2^{m+k+1}, &\tau=0, \textbf{t}_1=\textbf{t}'_1,\textbf{t}_3=\textbf{t}'_3,\\
%0        , & 0<|\tau|<2^{m-p},\textbf{t}_1=\textbf{t}'_1,\textbf{t}_3=\textbf{t}'_3,\\
%0        , & |\tau|<2^{m-p}, \textbf{t}_1\neq\textbf{t}'_1,\textbf{t}_3=\textbf{t}'_3,\\
%0        , & |\tau|<2^{m-p}, \textbf{t}_1=\textbf{t}'_1,\textbf{t}_3\neq\textbf{t}'_3,\\
%0        , & |\tau|<2^{m-p}, \textbf{t}_1\neq\textbf{t}'_1,\textbf{t}_3\neq\textbf{t}'_3.
%\end{cases}
%\end{split}
%\end{equation}
%From (\ref{eq1}), (\ref{eq2}), (\ref{eqqq2}), (\ref{eq6}), and (\ref{feq1}), we have
\begin{equation}\label{mainresult1}
\begin{split}
\mathscr{C}&(\psi(S_t),\psi(S_{t'}))(\tau)\\
&=\sum_{\textbf{u}_{W'}=\textbf{e}}(-1)^{(\textbf{t}_2-\textbf{t}'_2)\cdot \textbf{e}} \mathcal{A}\\
&=\begin{cases}
  2^{m-p+1}\displaystyle\sum_{\textbf{u}_{W'}=\textbf{e}}(-1)^{(\textbf{t}_2-\textbf{t}'_2)\cdot \textbf{e}}\sum_{\textbf{c}\in\{0,1\}^k}(-1)^{(\textbf{t}_1-\textbf{t}'_1)\cdot\textbf{c}}\\
  ~~~~~\times\displaystyle\sum_{\textbf{d}'\in\{0,1\}^p}
  (-1)^{(\textbf{t}_3-\textbf{t}_3')\cdot \textbf{d}'},~\tau=0\\
  2^{m-p+1}\displaystyle\sum_{\textbf{u}_{W'}=\textbf{e}}(-1)^{(\textbf{t}_2-\textbf{t}'_2)\cdot \textbf{e}}\\
  \times
  \displaystyle\sum_{(\textbf{d}',\textbf{d}'')\in \mathcal{K}_i}(-1)^{(\textbf{t}_3\cdot\textbf{d}'-\textbf{t}'_3\cdot\textbf{d}'')}\omega^{(\textbf{d}'-\textbf{d}'')\cdot (g_{m-p},\hdots,g_{m-1})}
  \\
\times\displaystyle\sum_{\textbf{c}\in\{0,1\}^k}(-1)^{(\textbf{t}_1-\textbf{t}'_1)\cdot\textbf{c}}\omega^{\mathcal{G}_{\textbf{c}\textbf{d}'}-\mathcal{G}_{\textbf{c}\textbf{d}''}},
\\
~~~~~~~~~~~~~~~~~~~~~~~~~~~~~~~~~~~~~\tau=\tau_i, i=1,2,\hdots,\varsigma\\
0, ~~~~~~~~~~~~~~~~~~~~~~~~~~~~~~~~~~~\textnormal{otherwise.}
 \end{cases}
\end{split}
\end{equation}
Similarly, we can show that
\begin{equation}\label{mainresult2}
\begin{split}
\mathscr{C}&(\psi^*(\bar{S}_t),\psi^*(\bar{S}_{t'}))(\tau)\\
&=\sum_{\textbf{u}_{W'}=\textbf{e}}(-1)^{(\textbf{t}_2-\textbf{t}'_2)\cdot \textbf{e}} \mathcal{A}\\
&=\begin{cases}
  2^{m-p+1}\displaystyle\sum_{\textbf{u}_{W'}=\textbf{e}}(-1)^{(\textbf{t}_2-\textbf{t}'_2)\cdot \textbf{e}}\sum_{\textbf{c}\in\{0,1\}^k}(-1)^{(\textbf{t}_1-\textbf{t}'_1)\cdot\textbf{c}}\\
  ~~~~~\times\displaystyle\sum_{\textbf{d}'\in\{0,1\}^p}
  (-1)^{(\textbf{t}_3-\textbf{t}_3')\cdot \textbf{d}'},~\tau=0\\
  2^{m-p+1}\displaystyle\sum_{\textbf{u}_{W'}=\textbf{e}}(-1)^{(\textbf{t}_2-\textbf{t}'_2)\cdot \textbf{e}}\\
  \times
  \displaystyle\sum_{(\textbf{d}',\textbf{d}'')\in \mathcal{K}_i}(-1)^{(\textbf{t}_3\cdot\textbf{d}'-\textbf{t}'_3\cdot\textbf{d}'')}\omega^{(\textbf{d}'-\textbf{d}'')\cdot (g_{m-p},\hdots,g_{m-1})}
  \\
\times\displaystyle\sum_{\textbf{c}\in\{0,1\}^k}(-1)^{(\textbf{t}_1-\textbf{t}'_1)\cdot\textbf{c}}\omega^{\mathcal{G}_{\textbf{c}\textbf{d}'}-\mathcal{G}_{\textbf{c}\textbf{d}''}},
\\
~~~~~~~~~~~~~~~~~~~~~~~~~~~~~~~~~~~~~\tau=\tau_i, i=1,2,\hdots,\varsigma\\
0, ~~~~~~~~~~~~~~~~~~~~~~~~~~~~~~~~~~~\textnormal{otherwise.}
 \end{cases}
\end{split}
\end{equation}
Finally, we need to find out $\mathscr{C}(\psi(S_t),\psi^*(\bar{S}_t))(\tau)$ for all $\tau$. By using (\ref{peq1}), $\mathscr{C}(\psi(S_t),\psi^*(\bar{S}_t))(\tau)$
can be expressed as 
\begin{equation}\label{tteq1}
\begin{split}
&\mathscr{C}(\psi(S_t),\psi^*(\bar{S}_{t'}))(\tau)\\
&=\displaystyle\sum_{\textbf{u}\in\{0,1\}^n}\mathscr{C}\left(g+\eta_1+\frac{q}{2}\left(\textbf{t}_1\cdot\textbf{x}_{J_1}+\textbf{t}_2\cdot \textbf{u}_{W'}+\textbf{t}_3\cdot\textbf{x}_\mathcal{Z}\right),\right.\\
&~~~~~~~~~~~~~\left.\tilde{g}+\eta_2+\frac{q}{2}\left(\textbf{t}'_1\cdot\bar{\textbf{x}}_{J_1}+\textbf{t}'_2\cdot \textbf{u}_{W'}+\textbf{t}'_3\cdot\bar{\textbf{x}}_\mathcal{Z}\right)\right)(\tau)\\
&=\sum_{\textbf{u}_{W'}=\textbf{e}}(-1)^{(\textbf{t}_2+\textbf{t}'_2)\cdot \textbf{e}}\\ &
\times\left( \sum_{\textbf{u}_W,u_{w_k}}\mathscr{C}\left(f\arrowvert_{\textbf{u}_{W'}=\textbf{e}}+h\arrowvert_{\textbf{u}_{W'}=\textbf{e}}+\eta_1+\frac{q}{2}\left(\textbf{t}_1\cdot\textbf{x}_{J_1}+\textbf{t}_3\cdot\textbf{x}_\mathcal{Z}\right),\right.\right.\\ 
&~~~\left.\left.\tilde{f}^*\arrowvert_{\textbf{u}_{W'}=\textbf{e}}+\eta_2+h^*\arrowvert_{\textbf{u}_{W'}=\textbf{e}}+\frac{q}{2}\left(\textbf{t}'_1\cdot\bar{\textbf{x}}_{J_1}+\textbf{t}'_3\cdot\bar{\textbf{x}}_\mathcal{Z}\right)\right)(\tau)\right)\\
&=\sum_{\textbf{u}_{W'}=\textbf{e}}(-1)^{(\textbf{t}_2-\textbf{t}'_2)\cdot \textbf{e}} \mathcal{A}_3,
\end{split}
\end{equation}
where
\begin{equation}
 \begin{split}
  \mathcal{A}_3&\\=&\sum_{\textbf{u}_W,u_{w_k}}\mathscr{C}\left(f\arrowvert_{\textbf{u}_{W'}=\textbf{e}}+h\arrowvert_{\textbf{u}_{W'}=\textbf{e}}+\eta_1+\frac{q}{2}\left(\textbf{t}_1\cdot\textbf{x}_{J_1}+\textbf{t}_3\cdot\textbf{x}_\mathcal{Z}\right),\right.\\ 
&~~~\left.\tilde{f}^*\arrowvert_{\textbf{u}_{W'}=\textbf{e}}+\eta_2+h^*\arrowvert_{\textbf{u}_{W'}=\textbf{e}}+\frac{q}{2}\left(\textbf{t}'_1\cdot\bar{\textbf{x}}_{J_1}+\textbf{t}'_3\cdot\bar{\textbf{x}}_\mathcal{Z}\right)\right)(\tau)\\
=& \sum_{\textbf{u}_Wu_{w_k}=\textbf{b}b}\mathscr{C}\left(f\arrowvert_{\textbf{u}_{W'},\textbf{u}_Wu_{w_k}=\textbf{e},\textbf{b}b}+h\arrowvert_{\textbf{u}_{W'},\textbf{u}_Wu_{w_k}=\textbf{e},\textbf{b}b}\right.\\&~~~~+\eta_1\arrowvert_{\textbf{u}_Wu_{w_k}=\textbf{b}b}+\frac{q}{2}\left(\textbf{t}_1\cdot\textbf{x}_{J_1}+\textbf{t}_3\cdot\textbf{x}_\mathcal{Z}\right),\\ 
&~~~~~\tilde{f}^*\arrowvert_{\textbf{u}_{W'},\textbf{u}_Wu_{w_k}=\textbf{e},\textbf{b}b}+h^*\arrowvert_{\textbf{u}_{W'},\textbf{u}_Wu_{w_k}=\textbf{e},\textbf{b}b}\\&~~~~\left.+\eta_2\arrowvert_{\textbf{u}_Wu_{w_k}=\textbf{b}b}+\frac{q}{2}\left(\textbf{t}'_1\cdot\bar{\textbf{x}}_{J_1}+\textbf{t}'_3\cdot\bar{\textbf{x}}_\mathcal{Z}\right)\right)(\tau)\\
\end{split}
\end{equation}
\begin{equation}\nonumber
\begin{split}
=& \sum_{\textbf{u}_Wu_{w_k}=\textbf{b}b}\omega^{2h\arrowvert_{\textbf{u}_{W'},\textbf{u}_Wu_{w_k}=\textbf{e},\textbf{b}b}}\mathscr{C}\left(f\arrowvert_{\textbf{u}_{W'},\textbf{u}_Wu_{w_k}=\textbf{e},\textbf{b}b}\right.\\&~~~~~+\eta_1\arrowvert_{\textbf{u}_Wu_{w_k}=\textbf{b}b}+\frac{q}{2}\left(\textbf{t}_1\cdot\textbf{x}_{J_1}+\textbf{t}_3\cdot\textbf{x}_\mathcal{Z}\right),\\ 
&~~~~~~\tilde{f}^*\arrowvert_{\textbf{u}_{W'},\textbf{u}_Wu_{w_k}=\textbf{e},\textbf{b}b}\\&~~~~~\left.+\eta_2\arrowvert_{\textbf{u}_Wu_{w_k}=\textbf{b}b}+\frac{q}{2}\left(\textbf{t}'_1\cdot\bar{\textbf{x}}_{J_1}+\textbf{t}'_3\cdot\bar{\textbf{x}}_\mathcal{Z}\right)\right)(\tau).
\end{split}
\end{equation}
From (\ref{cond}) and (\ref{tteq1}), we have 
\begin{equation}\label{tteq2}
 \begin{split}
  \mathcal{A}_3
&=\omega^{2\delta}\sum_{\textbf{u}_Wu_{w_k}=\textbf{b}b}\mathscr{C}\left(f\arrowvert_{\textbf{u}_{W'},\textbf{u}_Wu_{w_k}=\textbf{e},\textbf{b}b}\right.\\&~+\eta_1\arrowvert_{\textbf{u}_Wu_{w_k}=\textbf{b}b}+\frac{q}{2}\left(\textbf{t}_1\cdot\textbf{x}_{J_1}+\textbf{t}_3\cdot\textbf{x}_\mathcal{Z}\right),\\ 
&~~~~~~~~~~~\tilde{f}^*\arrowvert_{\textbf{u}_{W'},\textbf{u}_Wu_{w_k}=\textbf{e},\textbf{b}b}\\&~\left.+\eta_2\arrowvert_{\textbf{u}_Wu_{w_k}=\textbf{b}b}+\frac{q}{2}\left(\textbf{t}'_1\cdot\bar{\textbf{x}}_{J_1}+\textbf{t}'_3\cdot\bar{\textbf{x}}_\mathcal{Z}\right)\right)(\tau)\\
&=\omega^{2\delta}\mathcal{A}_4,
\end{split}
\end{equation}
where, 
\begin{equation}\label{tteq3}
 \begin{split}
  \mathcal{A}_4
&=\sum_{\textbf{u}_Wu_{w_k}=\textbf{b}b}\mathscr{C}\left(f\arrowvert_{\textbf{u}_{W'},\textbf{u}_Wu_{w_k}=\textbf{e},\textbf{b}b}\right.\\&~~~~~~~~~~~~+\eta_1\arrowvert_{\textbf{u}_Wu_{w_k}=\textbf{b}b}+\frac{q}{2}\left(\textbf{t}_1\cdot\textbf{x}_{J_1}+\textbf{t}_3\cdot\textbf{x}_\mathcal{Z}\right),\\ 
&~~~~~~~~~~~\tilde{f}^*\arrowvert_{\textbf{u}_{W'},\textbf{u}_Wu_{w_k}=\textbf{e},\textbf{b}b}\\&~~~~~~~~~~~~~\left.+\eta_2\arrowvert_{\textbf{u}_Wu_{w_k}=\textbf{b}b}+\frac{q}{2}\left(\textbf{t}'_1\cdot\bar{\textbf{x}}_{J_1}+\textbf{t}'_3\cdot\bar{\textbf{x}}_\mathcal{Z}\right)\right)(\tau).
\end{split}
\end{equation}
Assume, $\mathcal{F}_1=f\arrowvert_{\textbf{u}_{W'},\textbf{u}_Wu_{w_k}=\textbf{e},\textbf{b}b}$. From, (\ref{peq1}), we have
\begin{equation}\label{tteq4}
\begin{split}
\eta_1\arrowvert_{\textbf{u}_Wu_{w_k}=\textbf{b}b}
      &=\frac{q}{2}\left(\textbf{b}\cdot \textbf{x}_{J_1}+bx_{\gamma_{\textbf{c}}}\right),\\
\eta_2\arrowvert_{\textbf{u}_Wu_{w_k}=\textbf{b}b}
      &=\frac{q}{2}\left(\textbf{b}\cdot\bar{\textbf{x}}_{J_1}+\bar{b}x_{\gamma_\textbf{c}}\right).
\end{split}
\end{equation}
From (\ref{tteq4}), (\ref{tteq3}), and substituting $\mathcal{F}_1=f\arrowvert_{\textbf{u}_{W'},\textbf{u}_Wu_{w_k}=\textbf{e},\textbf{b}b}$ in 
(\ref{tteq3}), we have
\begin{equation}\label{tteq5}
 \begin{split}
  \mathcal{A}_4
&=\sum_{\textbf{u}_Wu_{w_k}=\textbf{b}b}\mathscr{C}\left(\mathcal{F}_1+\frac{q}{2}\left(\textbf{b}\cdot \textbf{x}_{J_1}+bx_{\gamma_{\textbf{c}}}\right)\right.\\&~~~~~~~~+\frac{q}{2}\left(\textbf{t}_1\cdot\textbf{x}_{J_1}+\textbf{t}_3\cdot\textbf{x}_\mathcal{Z}\right),\\ 
&~~~~~~~~~~~~~~~\tilde{\mathcal{F}}^*_1+\frac{q}{2}\left(\textbf{b}\cdot\bar{\textbf{x}}_{J_1}+\bar{b}x_{\gamma_\textbf{c}}\right)\\&~~~~~~~~~~~~~~~~~~~~~~\left.+\frac{q}{2}\left(\textbf{t}'_1\cdot\bar{\textbf{x}}_{J_1}+\textbf{t}'_3\cdot\bar{\textbf{x}}_\mathcal{Z}\right)\right)(\tau).
\end{split}
\end{equation}
For each $\textbf{c}\in \{0,1\}^k$, $G\left(\mathcal{F}_1\arrowvert_{\textbf{x}_{J_1}=\textbf{c}}\right)$ contains a path over the vertices 
specified in $J_1'$ and $p$ isolated vertices labeled $m-p,m-p+1,\hdots,m-1$. Therefore, by employing \textit{Lemma} \ref{lm1} in (\ref{tteq5}),
we have 
\begin{equation}\label{tteq6}
 \begin{split}
  \mathcal{A}_4
&=\sum_{\textbf{u}_Wu_{w_k}=\textbf{b}b}\mathscr{C}\left(\mathcal{F}_1+\frac{q}{2}\left(\textbf{b}\cdot \textbf{x}_{J_1}+bx_{\gamma_{\textbf{c}}}\right)\right.\\&~~~~+\frac{q}{2}\left(\textbf{t}_1\cdot\textbf{x}_{J_1}+\textbf{t}_3\cdot\textbf{x}_\mathcal{Z}\right),\\ 
&~~~~~~~~~~~~~~~\tilde{\mathcal{F}}^*_1+\frac{q}{2}\left(\textbf{b}\cdot\bar{\textbf{x}}_{J_1}+\bar{b}x_{\gamma_\textbf{c}}\right)\\&~~~~~~~~~~~~~~~~~~~~~~\left.+\frac{q}{2}\left(\textbf{t}'_1\cdot\bar{\textbf{x}}_{J_1}+\textbf{t}'_3\cdot\bar{\textbf{x}}_\mathcal{Z}\right)\right)(\tau)\\
&=0~\forall \tau.
\end{split}
\end{equation}
From, (\ref{tteq1}), (\ref{tteq2}), and (\ref{tteq6}), we have 
\begin{equation}\label{mainresult3}
 \mathscr{C}(\psi(S_t),\psi^*(\bar{S}_{t'}))(\tau)=0~\forall\tau.
\end{equation}
We have defined before $$\mathcal{K}_i\!=\!\left\{(\textbf{d}',\textbf{d}'')\!\!:\!\! \textbf{d}'\!\neq\! \textbf{d}'', (\textbf{d}'-\textbf{d}'')\!\cdot\! (2^{m-p},\hdots,2^{m-1})\!=\!\tau_i \right\},$$ for 
$i=1,2,\hdots,\varsigma$. Now, we shall find out $\displaystyle\min_{i\in\{1,2,\hdots,\varsigma\}} |\tau_i|$.
\begin{equation}\label{boun}
 \begin{split}
  |\tau_i|&=\left|(\textbf{d}'-\textbf{d}'')\!\cdot\! (2^{m-p},\hdots,2^{m-1})\right|\\
          &=\left|(d_1'-d_1'')2^{m-p}+\hdots+(d_p'-d_p'')2^{m-1}\right|\\
          &=2^{m-p}\left|(d_1'-d_1'')+\hdots+(d_p'-d_p'')2^{p-1}\right|\\
          &\geq 2^{m-p}.
 \end{split}
\end{equation}
In (\ref{boun}), the equality occurs if we take $\textbf{d}'=(1,0,\hdots,0)$ and $\textbf{d}''=(0,0,\hdots,0)$. There can exist another 
$\textbf{d}'$ and $\textbf{d}''$ for which the equality can also occur. Therefore,
\begin{equation}\label{boun1}
 \displaystyle\min_{i\in\{1,2,\hdots,\varsigma\}}\left|\tau_i\right|=2^{m-p}.
\end{equation}
From (\ref{mainresult1}), we have
\begin{equation}\label{mmres1}
\begin{split}
 \mathscr{C}(\psi(S_t),\psi(S_{t'}))(\tau)=\begin{cases}
                                  2^{m+n}, &\tau=0, t=t',\\
                                  0, & 0<|\tau|<2^{m-p}, t=t',\\
                                  0,&|\tau|<2^{m-p},t\neq t'.
                                 \end{cases}
 \end{split}
\end{equation}
From (\ref{mainresult2}), we have
\begin{equation}\label{mmres2}
\begin{split}
 \mathscr{C}(\psi(\bar{S}_t),\psi(\bar{S}_{t'}))(\tau)=\begin{cases}
                                  2^{m+n}, &\tau=0, t=t',\\
                                  0, & 0<|\tau|<2^{m-p}, t=t',\\
                                  0,&|\tau|<2^{m-p},t\neq t'.
                                 \end{cases}
 \end{split}
\end{equation}
Finally, from (\ref{mainresult3}), (\ref{mmres1}), and (\ref{mmres2}), we have
\begin{equation}
 \{\psi({S}_t): 0\leq t\leq 2^{n+p-1}-1\}\cup \{\psi^*(\bar{S}_t): 0\leq t\leq 2^{n+p-1}-1\},
 \end{equation}
is a $(2^{n+p},2^{m-p})$-$\text{ZCCS}_{2^{n}}^{2^m}$.

%Now,
%\begin{equation}\label{sres}
%\begin{split}
%\displaystyle\sum_{\textbf{u}_{W'}=\textbf{e}}(-1)^{(\textbf{t}_2-\textbf{t}'_2)\cdot \textbf{e}}
%=\begin{cases}
%2^{n-k-1}, & \textbf{t}_2=\textbf{t}'_2,\\
%0, & \textbf{t}_2\neq\textbf{t}'_2.
%\end{cases}
%\end{split}
%\end{equation}
%From (\ref{feq2}), and (\ref{sres}), we have
%\begin{equation}\label{feq3}
%\begin{split}
%C(\psi(S_t),\psi(S_{t'}))(\tau)
%&=\begin{cases}
%2^{m+n},&\tau=0, t=t',\\
%0        , &0<|\tau|<2^{m-p},t=t',\\
%0        ,&|\tau|<2^{m-p},t\neq t'.
%\end{cases}
%\end{split}
%\end{equation}
 \section{Proof of \textnormal{\textit{Theorem} \ref{thm2}}}
The $\psi(S_t)$ and $\psi(S_{t'})$ will be in a same code group if $\textbf{t}_1=\textbf{t}_1'$, $\textbf{t}_2=\textbf{t}_2'$, otherwise
the codes will be in two different code groups. The term $\mathcal{G}_{\textbf{x}_{J_1}\textbf{x}_\mathcal{Z}}$ is assumed to be zero in 
\textit{Theorem} \ref{thm2}. Therefore, by replacing 
$\mathcal{G}_{\textbf{c}\textbf{d}'}=\mathcal{G}_{\textbf{c}\textbf{d}''}
=0$ in (\ref{mainresult1}) and from (\ref{boun1}), we have
 \begin{equation}\label{igcmain1}
\begin{split}
\mathscr{C}&(\psi(S_t),\psi(S_{t'}))(\tau)\\
&=\begin{cases}
  2^{m-p+1}\displaystyle\sum_{\textbf{u}_{W'}=\textbf{e}}(-1)^{(\textbf{t}_2-\textbf{t}'_2)\cdot \textbf{e}}\sum_{\textbf{c}\in\{0,1\}^k}(-1)^{(\textbf{t}_1-\textbf{t}'_1)\cdot\textbf{c}}\\
  ~~~~~\times\displaystyle\sum_{\textbf{d}'\in\{0,1\}^p}
  (-1)^{(\textbf{t}_3-\textbf{t}_3')\cdot \textbf{d}'},~\tau=0\\
  2^{m-p+1}\displaystyle\sum_{\textbf{u}_{W'}=\textbf{e}}(-1)^{(\textbf{t}_2-\textbf{t}'_2)\cdot \textbf{e}}\\
  \times
  \displaystyle\sum_{(\textbf{d}',\textbf{d}'')\in \mathcal{K}_i}(-1)^{(\textbf{t}_3\cdot\textbf{d}'-\textbf{t}'_3\cdot\textbf{d}'')}\omega^{(\textbf{d}'-\textbf{d}'')\cdot (g_{m-p},\hdots,g_{m-1})}
  \\
\times\displaystyle\sum_{\textbf{c}\in\{0,1\}^k}(-1)^{(\textbf{t}_1-\textbf{t}'_1)\cdot\textbf{c}},
\\
~~~~~~~~~~~~~~~~~~~~~~~~~~~~~~~~~~~~~\tau=\tau_i, i=1,2,\hdots,\varsigma\\
0, ~~~~~~~~~~~~~~~~~~~~~~~~~~~~~~~~~~~\textnormal{otherwise.}
 \end{cases}\\
&=\begin{cases}
  2^{m+n},~\tau=0,~\textbf{t}_1=\textbf{t}_1',\textbf{t}_2=\textbf{t}_2',\textbf{t}_3=\textbf{t}_3',\\
  0,~0<|\tau|<2^{m-p}, ~\textbf{t}_1=\textbf{t}_1',\textbf{t}_2=\textbf{t}_2',\textbf{t}_3=\textbf{t}_3',\\
  0,~|\tau|<2^{m-p}, ~\textbf{t}_1=\textbf{t}_1',\textbf{t}_2=\textbf{t}_2',\textbf{t}_3\neq\textbf{t}_3',\\
  0,~|\tau|<2^m, ~\textbf{t}_1=\textbf{t}_1',\textbf{t}_2\neq\textbf{t}_2',\textbf{t}_3=\textbf{t}_3',\\
  0,~|\tau|<2^m, ~\textbf{t}_1=\textbf{t}_1',\textbf{t}_2\neq\textbf{t}_2',\textbf{t}_3\neq\textbf{t}_3',\\
  0,~|\tau|<2^m, ~\textbf{t}_1\neq\textbf{t}_1',\textbf{t}_2=\textbf{t}_2',\textbf{t}_3=\textbf{t}_3',\\
    0,~|\tau|<2^m, ~\textbf{t}_1\neq\textbf{t}_1',\textbf{t}_2=\textbf{t}_2',\textbf{t}_3\neq\textbf{t}_3',\\
      0,~|\tau|<2^m, ~\textbf{t}_1\neq\textbf{t}_1',\textbf{t}_2\neq\textbf{t}_2',\textbf{t}_3=\textbf{t}_3',\\
        0,~|\tau|<2^m, ~\textbf{t}_1\neq\textbf{t}_1',\textbf{t}_2\neq\textbf{t}_2',\textbf{t}_3\neq\textbf{t}_3'.
 \end{cases}
\end{split}
\end{equation}
Similarly, from (\ref{mainresult2}), we can show that
\begin{equation}\label{igcmain2}
\begin{split}
\mathscr{C}&(\psi^*(\bar{S}_t),\psi^*(\bar{S}_{t'}))(\tau)\\
&=\begin{cases}
  2^{m+n},~\tau=0,~\textbf{t}_1=\textbf{t}_1',\textbf{t}_2=\textbf{t}_2',\textbf{t}_3=\textbf{t}_3',\\
  0,~0<|\tau|<2^{m-p}, ~\textbf{t}_1=\textbf{t}_1',\textbf{t}_2=\textbf{t}_2',\textbf{t}_3=\textbf{t}_3',\\
  0,~|\tau|<2^{m-p}, ~\textbf{t}_1=\textbf{t}_1',\textbf{t}_2=\textbf{t}_2',\textbf{t}_3\neq\textbf{t}_3',\\
  0,~|\tau|<2^m, ~\textbf{t}_1=\textbf{t}_1',\textbf{t}_2\neq\textbf{t}_2',\textbf{t}_3=\textbf{t}_3',\\
  0,~|\tau|<2^m, ~\textbf{t}_1=\textbf{t}_1',\textbf{t}_2\neq\textbf{t}_2',\textbf{t}_3\neq\textbf{t}_3',\\
  0,~|\tau|<2^m, ~\textbf{t}_1\neq\textbf{t}_1',\textbf{t}_2=\textbf{t}_2',\textbf{t}_3=\textbf{t}_3',\\
    0,~|\tau|<2^m, ~\textbf{t}_1\neq\textbf{t}_1',\textbf{t}_2=\textbf{t}_2',\textbf{t}_3\neq\textbf{t}_3',\\
      0,~|\tau|<2^m, ~\textbf{t}_1\neq\textbf{t}_1',\textbf{t}_2\neq\textbf{t}_2',\textbf{t}_3=\textbf{t}_3',\\
        0,~|\tau|<2^m, ~\textbf{t}_1\neq\textbf{t}_1',\textbf{t}_2\neq\textbf{t}_2',\textbf{t}_3\neq\textbf{t}_3'.
 \end{cases}
\end{split}
\end{equation}
Also, from (\ref{mainresult3}), we have 
\begin{equation}\nonumber
 \mathscr{C}(\psi(S_t),\psi^*(\bar{S}_{t'}))(\tau)=0~\forall\tau.
\end{equation}
By using the results of (\ref{igcmain1}), (\ref{igcmain2}), and (\ref{mainresult3}), the code groups 
$\mathcal{I}_0,\mathcal{I}_1,\hdots,\mathcal{I}_{2^{n-1}-1},
\bar{\mathcal{I}}^*_{0},\bar{\mathcal{I}}^*_{1},\hdots,\bar{\mathcal{I}}^*_{2^{n-1}-1}$ forms an IGC code 
set $\mathcal{I}(2^{n+p},2^n,2^m,2^{m-p})$.
%The code $\psi(S_t)$ and $\psi(S_{t'})$ will be in a same code group if $\textbf{t}_1=\textbf{t}_1'$, $\textbf{t}_2=\textbf{t}_1'$, otherwise
%the codes will be in two different code groups. 
\end{appendices}
\bibliographystyle{IEEEtran}
\bibliography{notesk1}
\end{document}